\documentclass[letterpaper,twocolumn,10pt]{article}
\usepackage{usenix-2020-09}

\usepackage{tikz}
\usepackage{amsmath}

\usepackage{threeparttable}

\usepackage{graphicx}
\usepackage[many]{tcolorbox}
\usepackage{mathtools,latexsym,amsfonts,stmaryrd}
\usepackage{pbox}
\usepackage{amsthm}
\usepackage{multirow}
\usepackage{tikz}
\usepackage{mathrsfs}
\usepackage{mathpartir}
\usepackage[ruled,linesnumbered]{algorithm2e}
\usepackage{url}
\usepackage{syntax}
\usepackage{framed}
\usepackage[noend]{algpseudocode}
\usepackage{flushend}
\usepackage{listings}
\usepackage{moresize}
\usepackage{wrapfig}
 
\usepackage{seqsplit}
\usepackage{alltt}
\usepackage{xspace}
\usepackage{enumitem}
\usepackage{pifont}

\DeclareMathAlphabet{\mathcal}{OMS}{cmsy}{m}{n}

\usepackage{amsmath, listings, amsthm, proof, xspace}

\usepackage{thmtools}

\declaretheoremstyle[spaceabove=\topsep,notefont=\normalfont\itshape]{mystyle}

\newcommand{\revise}[2]{{\color{red}{\ifx&#1&\else- #1\fi}} {\color{ForestGreen}{\ifx&#2&\else+ #2\fi}}}\renewcommand{\revise}[2]{#2}

\usetikzlibrary{arrows,patterns, decorations.pathreplacing}

\newcommand{\F}{Fig.}

\newcommand{\T}{Table}
\renewcommand{\S}{Sec.}

\newcommand{\fixme}[1]{{#1}}
\newcommand{\mr}[1]{{#1}}

\newcommand{\ignore}[1]{}

\lstdefinestyle{base}{
  moredelim=**[is][\color{red}]{@}{@},
  escapeinside={<@}{@>}
}

\usepackage{pifont}

\newcommand\DejaVuttfamily{\fontfamily{DejaVuSansMono-TLF}\selectfont
}

\lstdefinestyle{base}{
  moredelim=**[is][\color{red}]{@}{@},
  escapeinside={<@}{@>}
}

\lstdefinelanguage
   [x64]{Assembler}     [x86masm]{Assembler} {morekeywords={CDQE,CQO,CMPSQ,CMPXCHG16B,JRCXZ,LODSQ,MOVSXD, POPFQ,PUSHFQ,SCASQ,STOSQ,IRETQ,RDTSCP,SWAPGS, rax,rdx,rcx,rbx,rsi,rdi,rsp,rbp, r8,r8d,r8w,r8b,r9,r9d,r9w,r9b,reg128,m128}} 

\let\OLDthebibliography\thebibliography
\renewcommand\thebibliography[1]{
  \OLDthebibliography{#1}
  \setlength{\parskip}{0pt}
  \setlength{\itemsep}{1pt plus 0.85ex}
}
\usepackage{listings}
\usepackage{fancyvrb}
\usepackage{color}
\definecolor{lightgray}{rgb}{.9,.9,.9}
\definecolor{darkgray}{rgb}{.4,.4,.4}
\definecolor{purple}{rgb}{0.65, 0.12, 0.82}
\definecolor{commentgreen}{RGB}{63,127,95}
\definecolor{pyblue}{RGB}{59,117,175}
\definecolor{pyorange}{RGB}{239,134,54}
\definecolor{pygreen}{RGB}{81,158,62}

\usepackage{xcolor}
\colorlet{myPurple}{blue!40!red}
\definecolor{myOrange}{RGB}{255,192,0}

\lstdefinelanguage{Solidity}{
  keywords={len,delete,int,void,payable, public, event, contract, typeof, new, true, false, catch, function, return, null, catch, switch, var, if, while, do, else, case, break,struct,const,socklen_t,sa_familty_t,char,sockaddr,load},
  keywordstyle=\color{violet}\bfseries,
  ndkeywords={class, export, boolean, throw, implements, import, this},
  ndkeywordstyle=\color{darkgray}\bfseries,
  identifierstyle=\color{black},
  sensitive=false,
  comment=[l]{//},
  escapeinside={(*@}{@*)},          morecomment=[s]{/*}{*/},
  commentstyle=\color{commentgreen}\ttfamily,
  stringstyle=\color{red}\ttfamily,
  morestring=[b]',
  morestring=[b]"
}

\newcommand{\rnum}[1]{\uppercase\expandafter{\romannumeral #1\relax}}

\usepackage{xcolor}

\algnewcommand{\LeftComment}[1]{\Statex \(\triangleright\) #1}

\definecolor{pptbrown}{RGB}{132,60,12}
\definecolor{pptgreen}{RGB}{56,87,35}
\definecolor{pptred}{RGB}{155,30,20}
\definecolor{pptdy}{RGB}{127,96,0}

\newcommand{\tool}{\textsc{BTD}\xspace}

\newcommand{\rom}[1]{\uppercase\expandafter{\romannumeral #1\relax}}

\newcommand*\circled[1]{\tikz[baseline=(char.base)]{
            \node[shape=circle,draw,inner sep=1.0pt] (char) {#1};}}

\begin{document}

\date{}

\title{\Large \bf Decompiling x86 Deep Neural Network Executables}
\author{
\rm Zhibo Liu, Yuanyuan Yuan, Shuai Wang\thanks{Corresponding author.} \\
The Hong Kong University of Science and Technology\\
\{zliudc,yyuanaq,shuaiw\}@cse.ust.hk
\and
{\rm Xiaofei Xie}\\
Singapore Management University \\
xfxie@smu.edu.sg
\and
{\rm Lei Ma}\\
University of Alberta \\
ma.lei@acm.org
}

\twocolumn
\maketitle

\begin{abstract}

  Due to their widespread use on heterogeneous hardware devices, deep learning
  (DL) models are compiled into executables by DL compilers to fully leverage
  low-level hardware primitives. This approach allows DL computations to be
  undertaken at low cost across a variety of computing platforms, including
  CPUs, GPUs, and various hardware accelerators.

  We present \tool\ (Bin to DNN), a decompiler for deep neural network (DNN)
  executables. \tool\ takes DNN executables and outputs full model
  specifications, including types of DNN operators, network topology,
  dimensions, and parameters that are (nearly) identical to those of the input
  models. \tool\ delivers a \mr{practical} framework to process DNN executables
  compiled by different DL compilers and with full optimizations enabled on x86
  platforms. It employs learning-based techniques to infer DNN operators,
  dynamic analysis to reveal network architectures, and symbolic execution to
  facilitate inferring dimensions and parameters of DNN operators.

  Our evaluation reveals that \tool\ enables accurate recovery of full
  specifications of complex DNNs with millions of parameters (e.g., ResNet). The
  recovered DNN specifications can be re-compiled into a new DNN executable
  exhibiting identical behavior to the input executable. We show that \tool\ can
  boost two representative attacks, adversarial example generation and knowledge
  stealing, against DNN executables. We also demonstrate cross-architecture legacy
  code reuse using \tool, and envision \tool\ being used for other critical
  downstream tasks like DNN security hardening and patching.

\end{abstract}

\section{Introduction}
\label{sec:introduction}

Recent years have witnessed increasing demand for applications of deep learning
(DL) in real-world scenarios. This demand has led to extensive deployment of DL
models in a wide spectrum of computing platforms, ranging from cloud servers to
embedded devices. Deployment of models in such a spread of platforms is
challenging, given the diversity of hardware characteristics involved (e.g.,
storage management and compute primitives) including GPUs, CPUs, and FPGAs.

A promising trend is to use DL compilers to manage and optimize these complex
deployments on multiple platforms~\cite{chen2018tvm,rotem2018glow,ma2020rammer}.
A DL compiler takes a high-level model specification (e.g., in ONNX
format~\cite{onnx}) and generates corresponding low-level optimized binary code
for a variety of hardware backends. For instance, TVM~\cite{chen2018tvm}, a
popular DL compiler, generates DNN executable with performance comparable to
manually optimized libraries; it can compile models for heterogeneous hardware
backends.
To date, DL compilers are already used by many edge devices and low-power chips
vendors~\cite{octoml,texas-instrument,qualcomm,nxp}. Cloud service
providers like Amazon and Google are also starting to use DL compiler in their
AI services for performance improvements~\cite{amazon,google}. In particular,
Amazon and Facebook are seen to spend considerable effort to compile DL models
on Intel x86 CPUs through the usage of DL
compilers~\cite{liu2019optimizing,jain2020efficient,facebook}.

Compilation of high-level models into binary code typically involves multiple
optimization cycles~\cite{chen2018tvm,rotem2018glow,ma2020rammer}. DL compilers
can optimize code utilizing domain-specific hardware features and abstractions.
Hence, generated executables manifest distinct representations of the high-level
models from which they were derived.
\mr{However, we observe that different low-level representations of the same DNN
operator in executables generally retain \textit{invariant} high-level
semantics, as DNN operators like ReLU and Sigmoid, are mathematically
defined in a rigorous manner. This reveals the opportunity of reliably recovering
high-level models by extracting semantics from each DNN operator's low-level
representation.}

\begin{figure*}[t]
  \centering
  \includegraphics[width=1.01\linewidth]{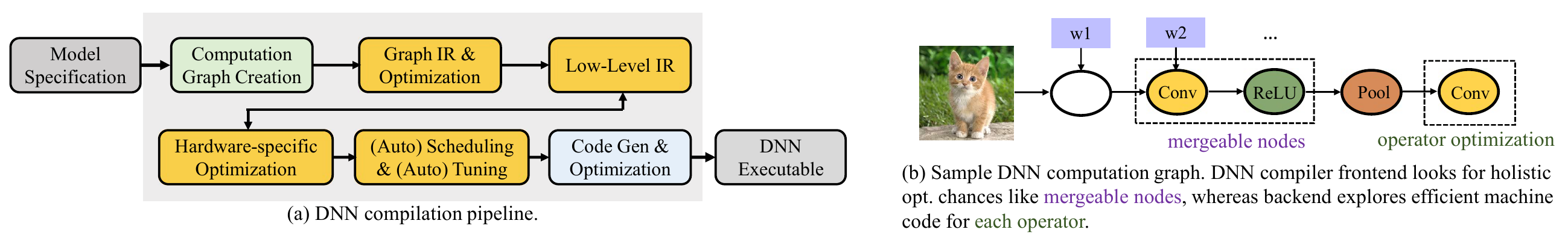}
  \vspace*{-25pt}
  \caption{The high-level workflow of DL compilation.}
  \vspace*{-10pt}
  \label{fig:dnn-compiler}
\end{figure*}

Extracting DNN models from executables can boost many security applications,
including adversarial example generation, training data inference, legacy DNN
model reuse, migration, and patching. In contrast, existing model-extraction
attacks, whether based on side
channels~\cite{hua2018reverse,duddu2018stealing,xiang2020open,hu2020deepsniffer,zhu2021hermes,yan2020cache}
or local
retraining~\cite{papernot2017practical,oh2019towards,teitelman2020stealing,orekondy2019knockoff},
assume specific attack environments or can leak only parts of DNN models with
low accuracy or high overhead.

We propose \tool, a decompiler for DNN executables. Given a (stripped)
executable compiled from a DNN model, we propose a three-step approach for full
recovery of DNN operators, network topology, dimensions, and parameters.
\tool\ conducts representation learning over disassembler-emitted assembly code
to classify assembly functions as DNN operators, such as convolution layers
(Conv).
Dynamic analysis is then used to chain DNN operators together, thus recovering
their topological connectivity. To further recover dimensions and parameters of
certain DNN operators (e.g., Conv), we launch trace-based symbolic
execution to generate symbolic constraints, primarily over
floating-point-related computations. \mr{The human-readable symbolic constraints
denote semantics of corresponding DNN operators that are invariant across
different compilation settings. Experienced DL experts can infer
higher-level information about operators (e.g., dimensions, the memory layout of
parameters) by reading the constraints. Nevertheless, to deliver an automated
pipeline, we then define patterns over symbolic constraints to
automatically recover dimensions and memory layouts of 
parameters}. We incorporate taint analysis to largely reduce the cost of
symbolic execution which is more heavyweight.

\mr{\tool\ is comprehensive as it handles \textit{all} DNN operators used in
forming computer vision (CV) models in ONNX Zoo~\cite{onnxzoo}.} \tool\
processes x86 executables, though its core technique is mostly
platform-independent. Decompiling executables on other architectures requires
vendor support for reverse engineering toolchains first. We also find that DNN
``executables'' on some other architectures are not in standalone executable
formats. See the last paragraph of \S~\ref{sec:preliminary} for the significance
of decompiling x86 DNN executables, and see \S~\ref{sec:discussion} for
discussions on cross-platform support.

\tool\ was evaluated by decompiling 64-bit x86 executables emitted by \mr{eight
versions of} three production DL compilers, TVM~\cite{chen2018tvm},
Glow~\cite{rotem2018glow}, NNFusion~\cite{ma2020rammer}, which are developed by
Amazon, Facebook, and Microsoft, respectively. These compilers enable full
optimizations during our evaluation.
\tool\ is scalable to recover DNN models from \mr{65} DNN executables, including
\mr{nearly 3 million} instructions, in \mr{60 hours} with negligible errors.
\tool, in particular, can recover over 100 million parameters from VGG, a large
DNN model, with an error rate of less than 0.1\% (for TVM-emitted executable)
or none (for Glow-emitted executable). Moreover, \mr{to demonstrate \tool's
correctness, we rebuild decompiled model specifications with PyTorch.
The results show that almost} all decompiled DNN models can be
recompiled into new executables that behave \textit{identically} to the
reference executables. We further demonstrate that \tool, by decompiling
executables into DNN models, can boost two attacks, adversarial example
generation and knowledge stealing. We also migrate decompiled x86 DNN
executables to GPUs, and discuss limits and potential future works. In summary,
we contribute the following:

\begin{itemize}[noitemsep,topsep=0pt,leftmargin=3mm] 
\item This paper, for the first time\footnote[1]{This paper was submitted to
  USENIX Security 2022 (Fall Round) on October 12, 2021. We received the Major
  Revision decision and re-submitted revised version to USENIX Security 2023
  (Summer Round) on June 07, 2022. When preparing the camera-ready version, we
  notice a parallel work DnD~\cite{wu2022dnd}, which considers decompiling DNN
  executables across architectures (\tool\ only considers x86 executables).
  Nevertheless, DnD does not deeply explore the impact of compiler optimizations
  compared to our work.}, advocates for reverse engineering DNN executables.
  \tool\ accepts as input (stripped) executables generated by production DL
  compilers and outputs complete model specifications. \tool\ can be used to aid
  in the comprehension, migration, hardening, and exploitation of DNN
  executables.

\item \tool\ features a three-step approach to recovering high-level DNN models.
  It incorporates various design principles and techniques to deliver an
  effective pipeline.

\item We evaluate \tool\ against executables compiled from large-scale DNN
  models using production DL compilers. \tool\ achieves high accuracy in
  recovering (nearly) full specifications of complex DNN models. We also
  demonstrate how common attacks are boosted by \tool.
\end{itemize}

 \section{Preliminary}
\label{sec:preliminary}

\F~\hyperref[fig:dnn-compiler]{1(a)} depicts DNN model compilation. DNN
compilation can be divided into two phases~\cite{li2020deep}, with each phase
manipulates one or several intermediate representations (IR).

\noindent \textbf{Computation Graph.}~DL compiler inputs are typically
high-level model descriptions exported from DL frameworks like
PyTorch~\cite{paszke2019pytorch}. DNN models are typically represented as
computation graphs in DL frameworks. \F~\hyperref[fig:dnn-compiler]{1(b)} shows
a simple graph of a multilayer convolutional neural network (CNN). These graphs
are usually high-level, with limited connections to hardware. DL frameworks
export computation graphs often in ONNX format~\cite{onnx} as DL compiler
inputs.

\noindent \textbf{Frontend: Graph IRs and Optimizations.}~DL compilers typically
first convert DNN computation graphs into graph IRs. Hardware-independent graph
IRs define graph structure. Network topology and layer dimensions encoded in
graph IRs can aid graph- and node-level optimizations including operator fusion,
static memory planning, and layout
transformation~\cite{chen2018tvm,rotem2018glow}. For instance, operator fusions
and constant folding are used to identify mergeable nodes in graph IRs after 
precomputing statically-determinable components.
Graph IRs specify high-level inputs and outputs of each operator, but do
not restrict how each operator is implemented.

\noindent \textbf{Backend: Low-Level IRs and Optimizations.}~Hardware-specific
low-level IRs are generated from graph IRs. Instead of translating graph IRs
directly into standard IRs like LLVM IR~\cite{Lattner2004LLVM}, low-level IRs
are employed as an intermediary step for customized optimizations using prior
knowledge of DL models and hardware characteristics. Graph IR operators can be
converted into low-level linear algebra operators~\cite{rotem2018glow}. For
example, a fully connected (FC) operator can be represented as matrix
multiplication followed by addition. Such representations alleviate the hurdles
of directly supporting many high-level operators on each hardware target.
Instead, translation to a new hardware target only needs the support of
low-level linear algebra operators. Low-level IRs are usually memory related.
Hence, optimizations at this step can include hardware intrinsic mapping, memory
allocation, loop-related optimizations, and
parallelization~\cite{ragan2013halide,
baghdadi2019tiramisu,yuki2012alphaz,chen2018tvm}.

\begin{figure*}[t]
  \centering
  \includegraphics[width=0.92\linewidth]{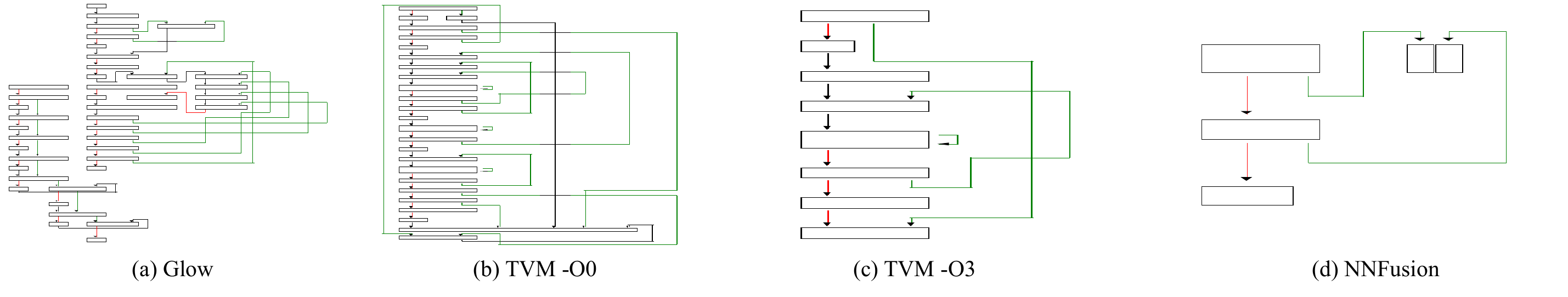}
  \vspace*{-10pt}
  \caption{Compare CFGs of a Conv operator in VGG16 compiled by different DL
    compilers. TVM refers to enabling no optimization as ``-O0'' while enabling
    full optimizations as ``-O3''. Glow and NNFusion by default apply full
    optimizations.}
  \vspace*{-10pt}
  \label{fig:motivation}
\end{figure*}

\noindent \textbf{Backend: Scheduling and Tuning.}~Policies mapping an operator
to low-level code are called \textit{schedules}. A compiler backend often
searches a vast combinatorial scheduling space for optimal parameter settings
like loop unrolling factors. Halide~\cite{ragan2013halide} introduces a
scheduling language with manual and automated schedule optimization primitives.
Recent works explore launching auto scheduling and tuning to enhance
optimization~\cite{adams2019learning, vasilache2018tensor,
mullapudi2016automatically, chen2018learning, chen2018tvm, zheng2020ansor,
zheng2020flextensor}. These methods alleviate manual efforts to decide schedules
and optimal parameters. 

\noindent \textbf{Backend: Code Gen.}~Low-level IRs are compiled to generate
code for different hardware targets like CPUs and GPUs. When generating machine
code, \textit{a DNN operator (or several fused operators) is typically compiled
into an individual assembly function}. Low-level IRs can be converted into
mature tool-chains IRs like LLVM or CUDA IR~\cite{nvidiair} to explore
hardware-specific optimizations. For instance, Glow~\cite{rotem2018glow} can
perform fine-grained loop-oriented optimizations in LLVM IR. DL compilers like
TVM and Glow compile optimized IR code into standalone executables. Kernel
libraries can be used by DL compilers NNFusion~\cite{ma2020rammer} and
XLA~\cite{xla} to statically link with DNN executables. Decompiling executables
statically linked with kernel libraries are \textit{much easier}: such
executables contain many wrappers toward kernel libraries. These wrappers (e.g.,
a trampoline to the Conv implementation in kernel libraries) can be used to
infer DNN models. This work mainly focuses on decompiling ``self-contained''
executables emitted by TVM and Glow, given their importance and difficulty. For
completeness, we demonstrate decompiling NNFusion-emitted executables in
\S~\ref{subsec:app-nnfusion}.

\noindent \textbf{Real-World Significance of DL Compilers.}~DL compilers offer
systematic optimization to improve DNN model adoption. Though many DNN models to
date are deployed using DL frameworks like Tensorflow, DL compilers cannot be
disregarded as a growing trend. Edge devices and low-power processors suppliers
are incorporating DL compilers into their applications to reap the benefits of
DNN models~\cite{octoml,texas-instrument,qualcomm,nxp}. Cloud service
providers like Amazon and Google include DL compilers into their DL services to
boost performance~\cite{amazon,google}. Amazon uses DL compilers to compile DNN
models on Intel x86 CPUs~\cite{liu2019optimizing,jain2020efficient}. Facebook
deploys Glow-compiled DNN models on Intel CPUs~\cite{facebook}. Overall, DL
compilers are increasingly vital to boost DL on Intel CPUs, embedded devices,
and other heterogeneous hardware backends. We design \tool, a decompiler for
Intel x86 DNN executables. We show how \tool\ can accelerate common DNN attacks
(\mr{Appendix~\ref{subsec:eval-attack}}) and migrate DNN executables to GPUs
(\S~\ref{sec:discussion}). \S~\ref{sec:discussion} explains why \tool\ does not
decompile executables on GPUs/accelerators. GPU/accelerator platforms lack
disassemblers/dynamic instrumentation infrastructures, and the DL compiler
support for GPU platforms is immature (e.g., cannot generate standalone
executables).
 
\section{Decompiling DNN Executables}
\label{sec:motivation}

\noindent \textbf{Definition.}~\tool\ decompiles DL executables to recover DNN
high-level specifications. The full specifications include: \circled{1} DNN
operators (e.g., ReLU, Pooling, and Conv) and their topological connectivity,
\circled{2} dimensions of each DNN operator, such as \#channels in Conv, and
\circled{3} parameters of each DNN operator, such as weights and biases, which
are important configurations learned during model training. \S~\ref{sec:design}
details \tool's processes to recover each component.

\noindent \textbf{Query-Based Model Extraction.}~Given a (remote) DNN model
with obscure specifications, adversaries can continuously feed inputs $x$
to the model and collect its prediction outputs $y$. This way, adversaries can
gradually assemble a training dataset $(x,y)$ to train a local
model~\cite{tramer2016stealing,papernot2017practical}.

This approach may have the following challenges: 1) for a DNN executable without
prior knowledge of its functionality, it is unclear how to prepare inputs $x$
aligned with its normal inputs; 2) even if the functionality is known, it may
still be challenging to prepare a non-trivial collection of $x$ for models
trained on private data (e.g., medical images); 3) local retraining may require
rich hardware and is costly; and 4) existing query-based model extraction
generally requires prior knowledge of model architectures and
dimensions~\cite{papernot2017practical}. \mr{In contrast, \tool only requires a
valid input. For instance, a meaningless image is sufficient to decompile
executables of CV models. Also}, according to the notation in
\textbf{Definition}, local retraining assumes \circled{1} + \circled{2} as prior
knowledge, whereas \tool\ fully recovers \circled{1} + \circled{2} + \circled{3}
from DNN executables.

\noindent \textbf{Model Extraction via Side Channels.}~Architectural-level hints
(e.g., side channels) leaked during model inference can be used for model
extraction~\cite{hua2018reverse,duddu2018stealing,xiang2020open,hu2020deepsniffer,zhu2021hermes,yan2020cache}.
These works primarily recover high-level model architecture, which are
\circled{1} or \circled{1} + \circled{2} according to our notation in
\textbf{Definition}. In contrast, \mr{\tool\ statically recovers \circled{1} and
then dynamically recovers \circled{2} + \circled{3} from DNN executables (but
coverage is not an issue; see \S~\ref{subsec:design-topology} for
clarification)}. \S~\ref{sec:related} further compares \tool\ with prior model
extraction works.

\noindent \textbf{Comparison with C/C++ Decompilation.}~\tool\ is
\textit{different} from C/C++ decompilers. C/C++ decompilation takes executable
and recovers C/C++ code that is visually similar to the original source code.
Contrarily, we explore decompiling DNN executables to recover original DNN
models. The main differences and common challenges are summarized below.

\noindent \underline{Statements vs.~Higher-Level Semantics:}~Software
decompilation, holistically speaking, line-by-line translates machine
instructions into C/C++ statements. In contrast, \mr{\tool\ recovers
higher-level model specifications from machine instructions}. This difference
clarifies that a C decompiler is \textit{not} \mr{sufficient} for decompilation
of DNN executables.

\noindent \mr{\underline{Common Uncertainty:}~There is no fixed mapping between
C/C++ statements and assembly instructions. Compilers may generate distinct
low-level code for the same source statements. Therefore, C/C++ decompilers
extensively use heuristics/patterns when mapping assembly code back to source
code. Likewise, DL compilers may adopt different optimizations for compiling the
same DNN operators. The compiled code may exhibit distinct syntactic forms.
Nevertheless, the semantics of DNN operators are retained, and we extract the
invariant semantics from the low-level instructions to infer the high-level
model specifications.
See \S~\ref{subsec:design-parameter} for details.}

\noindent \underline{End Goal:}~C/C++ compilation prunes high-level program
features, such as local variables, types, symbol tables, and high-level control
structures. Software decompilation is fundamentally
undecidable~\cite{cifuentes1995decompilation}, and to date, decompiled C/C++
code mainly aids (human-based) analysis and comprehension, \textit{not}
recompilation. Generating ``recompilable'' C code is very
challenging~\cite{wang2015uroboros, wang2017ramblr, bauman2018superset,
williams2020egalito}. In this regard, DNN compilation has \mr{comparable
difficulty, as} compilation and optimization discard information from DNN
\mr{models} (e.g., by fusing neighbor operators). \mr{\tool\ decompiles DNN
executables into high-level DNN specifications, resulting in a functional
executable after recompilation.} Besides helping (human-based) comprehension,
\tool\ boosts model reuse, migration, security hardening, and adversarial
attacks. See case studies in \S~\ref{sec:discussion} and
Appendix~\ref{subsec:eval-attack}.

\noindent \textbf{Opacity in DNN Executables.}~\F~\ref{fig:motivation} compares
VGG16~\cite{simonyan2014very} executables compiled using three DL compilers. 
For simplicity, we only plot the control flow graphs (CFGs) of
VGG16's first Conv operator. These CFGs were extracted using IDA-Pro~\cite{ida}.
Although this Conv is only one of 41 nodes in VGG16, Glow compiles it into a
dense CFG (\F~\hyperref[fig:motivation]{2(a)}). \S~\ref{sec:preliminary} has
introduced graph-level optimizations that selectively merge neighbor nodes.
Comparing CFG generated by TVM -O0 (\F~\hyperref[fig:motivation]{2(b)}) and
by TVM -O3 (\F~\hyperref[fig:motivation]{2(c)}), we find that
optimizations (e.g., operator fusion) in TVM can make CFG more succinct. We also
present CFGs emitted by NNFusion in \F~\hyperref[fig:motivation]{2(d)}:
NNFusion-emitted executables are coupled with the \texttt{Mlas}~\cite{mlas}
kernel library. This CFG depicts a simple trampoline to the Conv implementation
in \texttt{MlasGemm}.

As in \F~\ref{fig:motivation}, different compilers and optimizations can result
in complex and distinct machine code realizations. However, \tool\ is designed
as a \mr{general} approach for decompilation of executables compiled by these
diverse settings.

\noindent \textbf{Design Focus.}~Reverse engineering is generally sensitive to
the underlying platforms and compilation toolchains. \mr{As the first piece of
work in this field}, \tool\ is designed to process \textit{common} DNN models
compiled by standard DL compilers. Under such conservative and practical
settings, \tool\ delivers highly encouraging and accurate decompilation.
Similarly, obfuscation can impede C/C++ decompilation~\cite{liu2020far}. Modern
C/C++ decompilers are typically benchmarked on common software under standard
compilation and
optimization~\cite{williams2020egalito,brumley2013native,andriesse2016depth,wang2015uroboros},
instead of extreme cases. We leave it as a future work to study decompiling
obfuscated DL executables.

\begin{figure*}[t]
  \centering
  \includegraphics[width=1.00\linewidth]{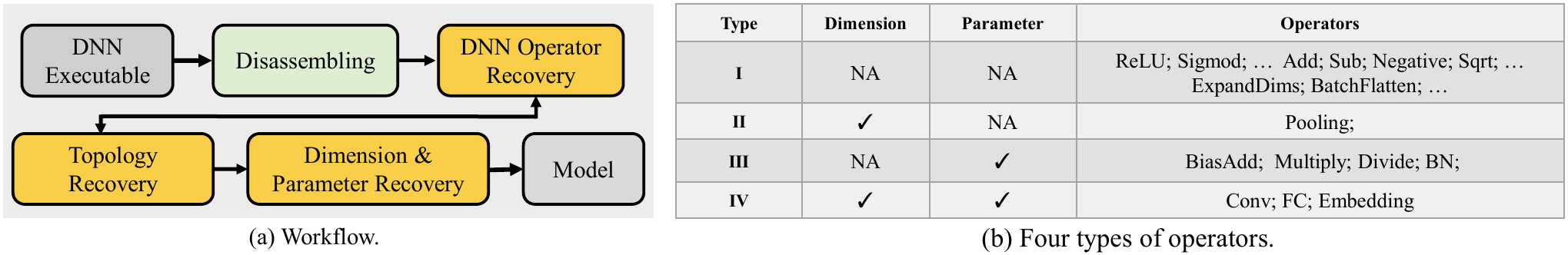}
  \vspace*{-25pt}
  \caption{Decompilation workflow. Here ``NA'' in the ``Dimension'' column
    denotes an easy case where output dimension of an operator \texttt{O} equals
    to its input dimension and no other dimensions associated with \texttt{O}.
    We find that in non-trivial DNN, it is sufficient to decide \texttt{O}'s
    dimensions after propagating dimensions from other operators on the DNN
    computation graph.}
  \vspace*{-10pt}
  \label{fig:workflow}
\end{figure*}
 
\section{Design}
\label{sec:design}

\mr{Decompiling DNN executables is challenging due to the mismatch between
instruction-level semantics and high-level model specifications. DNN executables
lack high-level information regarding operators, topologies, and dimensions.
Therefore, decompiling DNN executables presents numerous reverse engineering
hurdles, as it is difficult to deduce high-level model specifications from
low-level instructions. We advocate DL decompilers to satisfy the following
criteria:

\noindent \textbf{R1} (\textbf{Generalizability}): Avoid brittle assumptions. Generalize
across compilers, optimizations, and versions. 

\noindent \textbf{R2} (\textbf{Correctness}): Use effective, resilient methods and produce
correct outputs. 

\noindent \textbf{R3} (\textbf{Performance}): Be efficient when necessary.

\noindent \textbf{R4} (\textbf{Automation}): Avoid manual analysis and automate the
decompilation process.

\tool\ delivers practical decompilation based on the \textit{invariant
semantics} of DNN operators that aims to meet all four criteria. Our intuition
is simple: \textit{DL compilers generate distinct low-level code but retain
operator high-level semantics, because DNN operators are generally defined in a
clean and rigorous manner.} Therefore, recovering operator semantics should
facilitate decompilation generic across compilers and optimizations
(\textbf{R1}). Besides, as invariant semantics reflect high-level information,
e.g., operator types and dimensions, we can infer model abstractions accurately
(\textbf{R2}).}

\F~\hyperref[fig:workflow]{3(a)} depicts the BTD workflow.
\S~\ref{subsec:design-operator} describes \mr{learning-based} techniques for
recognizing assembly functions as DNN operators like Conv. Given recovered DNN
operators, we reconstruct the network topology using dynamic analysis
(\S~\ref{subsec:design-topology}). \mr{We then use trace-based symbolic
execution to extract operator semantics from assembly code and then recover
dimensions and parameters with semantics-based patterns (\S~\ref{subsubsec:se}).
Some operators are too costly for symbolic execution to analyze. We use taint
analysis to keep only tainted sub-traces for more expensive symbolic execution
to analyze (\textbf{R3}), as noted in \S~\ref{subsubsec:trace}. \tool\ is an
end-to-end, fully automated DNN decompiler (\textbf{R4}). \tool\ produces model
specifications that behave identically to original models, whose focus and
addressed challenges are distinct from C/C++ decompilation. \tool\ does not
guarantee 100\% correct outputs. In \S~\ref{sec:implementation}, we discuss
procedures users can follow to fix errors.}

\mr{Dimensions and parameters configure DNN operators}. We show representative
cases in \F~\hyperref[fig:workflow]{3(b)}. Type \rom{1} operators, including
activation functions like ReLU and \mr{element-wise arithmetic} operators, do
not ship with parameters; recovering their dimensions is trivial, as clarified
in the caption of \F~\ref{fig:workflow}. \mr{Type \rom{2} and \rom{3} operators
require dimensions or parameters, such as Pooling's stride $S$ and kernel size
$K$.} In addition to simple arithmetic operators, BiasAdd involves bias $B$, as
extra parameters. Type \rom{4} operators require both parameters and dimensions.
These operators form most DNN models. \S~\ref{subsec:comprehension} empirically
demonstrates ``comprehensivness'' of our study.

\mr{\tool\ recovers dimensions/parameters of all DNN operators used by CV models in
ONNX Zoo (see \S~\ref{subsec:comprehension}).} Due to limited space,
\S~\ref{subsec:design-parameter} only discusses decompiling \mr{the most
challenging operator,} Conv. The core techniques explained in
\S~\ref{subsec:design-parameter} are utilized to decompile other DNN operators.
However, \mr{other operators may use different (but simpler) patterns}.
Appendix~\ref{sec:recovery} lists other operator patterns. \mr{We further
discuss the extensibility of \tool in \S~\ref{subsec:exten}.}

\noindent \textbf{Disassembling and Function Recovery.}~\tool\ \mr{targets}
64-bit x86 executables. \mr{Cross-platform support is discussed} in
\S~\ref{sec:discussion}. \tool\ \mr{supports stripped executables without symbol
or debug information.}
We assume that DNN executables can be first flawlessly disassembled \mr{with
assembly functions recovered}. According to our observation, obstacles that can
undermine disassembly and function recovery in x86 executables, e.g.,
instruction overlapping and embedded data~\cite{bauman2018superset}, are
\textit{not} found in even highly-optimized DNN executables. We use a commercial
decompiler, IDA-Pro~\cite{ida} (ver.~7.5), to maximize confidence in the
credibility of our results.

\noindent \textbf{Compilation Provenance.}~Given a DNN executable $e$,
compilation provenance include: 1) which DL compiler is used, and 2) whether $e$
is compiled with full optimization -O3 or no optimization -O0. \mr{Since some
DNN operators (e.g., type IV in \F~\hyperref[fig:workflow]{3(b)}) in $e$ are
highly optimized when compiled, the compilation provenance can be inferred
\textit{automatically} by analyzing patterns over sequences of x86 instructions
derived from $e$. We extend our learning-based method from
\S~\ref{subsec:design-operator} to predict compilation provenance from assembly code.
Our evaluation of over all CV models in ONNX Zoo finds \textit{no} errors.}
Overall, we assume that compilation provenance is known to \tool. Therefore,
some patterns can be designed separately for Glow- and TVM-emitted executables;
see details in Appendix~\ref{sec:recovery}. To show $e$'s decompilation is flawless,
we \textit{must} recompile decompiled DNN models with the \textit{same
provenance} (see \S~\ref{subsec:eval-recompile}). \mr{Using
different compilation provenances may induce (small) numerical accuracy
discrepancies and is undesirable.}

\mr{This section focuses on decompilation of self-contained DNN executables
compiled by TVM and Glow. Decompilation of NNFusion-emitted executables is
easier because of its distinct code generation paradigm. We discuss decompiling
NNFusion-emitted executables in \S~\ref{subsec:app-nnfusion}.}

\subsection{DNN Operator Recovery}
\label{subsec:design-operator}

As introduced in \S~\ref{sec:preliminary}, one or a few \mr{fused} DNN operators
are compiled into an assembly function. We train a neural model to map assembly
functions to DNN operators. Recent works perform representation learning by
treating x86 opcodes as natural language tokens~\cite{ding2019asm2vec,
pei2020trex, yu2020order, duan2020deepbindiff, li2021palmtree}. These works help
comprehend x86 assembly code and assist downstream tasks like matching similar
code. Instead of defining explicit patterns over x86 opcodes to infer DNN
operators (which could be tedious and need manual efforts), we use
representation learning and treat x86 opcodes as language tokens.

\noindent \textbf{Atomic OPs.}~Launching representation learning directly over
x86 opcodes syntax can result in poor learning quality. Due to x86 instructions'
flexibility, opcodes with (nearly) identical semantics may have distinct
syntactic forms, e.g., \texttt{vmulps} and \texttt{mulps} denoting multiply over
floating numbers of different sizes. Rare words machine
translation~\cite{sennrich-2016-neural} are recently advanced from the
observation that natural language words can be divided into atomic units.
Translators can use atomic units to translate rare words. Accordingly, we define
\textit{atomic OPs} over x86 opcodes: an atomic OP represents an atomic and
indivisible unit of an x86 opcode. Each opcode is thus split into atomic OPs.
While a DNN operator could be compiled into various x86 opcode sequences,
induced atomic OP sequences can better reflect ``semantics'' in a
noisy-resilient way.

\noindent \textbf{Dividing Opcodes into Atomic OPs.}~As a common approach, we
segment opcodes using Byte Pair Encoding (BPE)~\cite{gage1994new}. BPE
iteratively replaces the most frequent consecutive bytes in a sequence with a
single, unused byte. We split each opcode into a sequence of characters and
counted consecutive characters to find the most frequent ones. BPE iterates
until the opcodes of all atomic OPs have been merged. For instance, opcodes
\texttt{vmulps} and \texttt{mulps} are first split into ``v m u l p s'' and ``m
u l p s'', and an atomic OP \texttt{mulps} is eventually extracted.

\noindent \textbf{Learning over Atomic OPs.}~We train a neural identifier model
with a sequence of atomic OPs from an assembly function as inputs. This model
outputs a 1D vector with $N$ dimensions ($N$ is the total number of unique DNN
operators), where multiple ``1'' in the vector implies that this assembly
function represents several fused DNN operators. All ``0'' in the vector implies
this function may be a DL compiler-inserted utility function (e.g., for memory
management). \mr{The order of fused operators is represented in symbolic
constraints extracted in \S~\ref{subsec:design-parameter}. Thus, predicted operator
labels and network topology will be refined after symbolic execution.}
Our model's frontend learns a neural embedding for each atomic OP and then
embeds a function's entire atomic OP sequence. The order of atomic OPs is found
to be vital in prediction. Therefore, we preserve the order of collected atomic
OPs within the assembly function. We encode orders with
LSTM~\cite{hochreiter1997long} (see \S~\ref{sec:implementation}) and enhance
learning with neural attention~\cite{bahdanau2014neural}.

\mr{\noindent \textbf{From Operators to Compilation Provenance.}~As noted in
\S~\ref{sec:design}, our decompilation pipeline requires compilation provenance.
We extend the model presented in this section to recover compilation provenance.
The extended model predicts compilation provenance using embeddings of all
functions in an executable as its input (function embeddings have been generated
above). We clarify this task is generally simple; humans can easily distinguish
assembly functions in executables from different compilation provenances. }
 \begin{figure*}[t]
  \centering
  \includegraphics[width=0.90\linewidth]{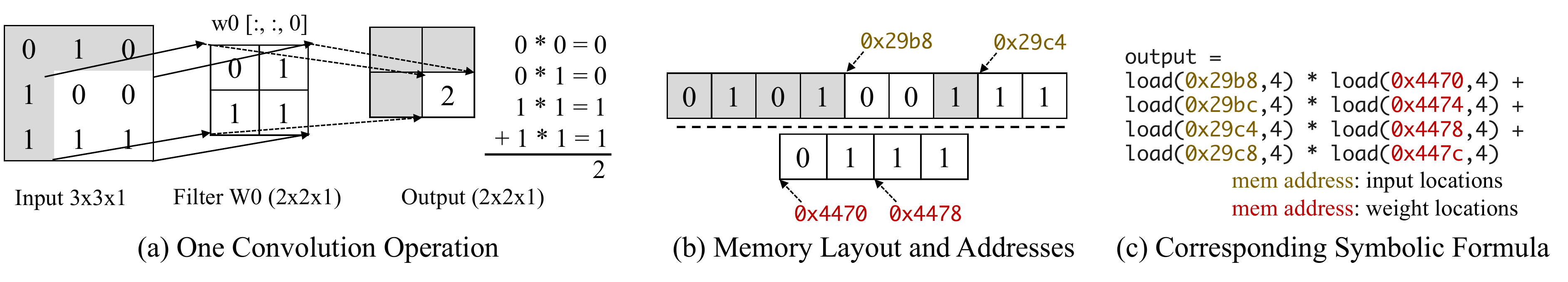}
  \vspace*{-10pt}
  \caption{Launching trace-based symbolic execution (SE) to infer dimensions and
    localize parameters for Conv operators.}
  \vspace*{-10pt}
  \label{fig:conv-se}
\end{figure*}

\subsection{DNN Network Topology Recovery}
\label{subsec:design-topology}

\mr{Recovering DNN network topology is straightforward, regardless of underlying
operator semantics. DNN operators are chained into a computation graph.
Generally, a DNN operator has a fixed number of inputs and
outputs~\cite{onnxop}. According to our observation, DL compilers compile DNN
operators into assembly functions and pass inputs and outputs as memory pointers
through function arguments. We use Intel Pin~\cite{pin}, a dynamic
instrumentation tool, to hook every callsite. During runtime, we record the
memory addresses of inputs/outputs passed to callsites and connect two operators
if the successor's inputs match the predecessor's outputs.

We do not rely on any compiler-specific assumptions like function signatures.
This step is independent of later steps and only uses shallow information
readily available in binaries. In case the inputs and outputs are passed
differently (e.g., not using pointers) in further compiler implementation
changes, we envision updating the instrumented code accordingly without much
engineering effort required. Also, we clarify that this dynamic analysis is not
limited by ``coverage''. We do not require ``semantically meaningful'' inputs
(not like a query-based model
extraction~\cite{tramer2016stealing,papernot2017practical}), just a format-valid
input to record how each operator in the executable accesses memory. One
format-valid, trivial (meaningless) input can achieve 100\% coverage. Besides,
whether a model accepts images or text as its valid inputs is easy to
determine.}

 \subsection{Dimension and Parameter Recovery}
\label{subsec:design-parameter}

As in \F~\ref{fig:workflow}, certain complex DNN operators are configured with
dimensions and parameters. \mr{This section details solutions to recover
parameters/dimensions. To present a comprehensive working example, we use Conv,
the most complex operator in our dataset, to introduce our solutions.
Nonetheless, our solutions are general enough to cover all operators in CV
models of ONNX Zoo (see \S~\ref{subsec:comprehension}) and can be extended to
support more operators with trivial effort (see \S~\ref{subsec:exten}).}

\F~\hyperref[fig:conv-se]{4(a)} shows the input, the kernel, and the output of a
simple Conv computation. Suppose no optimization is applied, we present the
memory layout of Conv in \F~\hyperref[fig:conv-se]{4(b)}. Inputs and parameters
are typically stored \textit{separately} in memory, whereas neighbor
input/parameter elements are stored \textit{contiguously}.
\mr{\F~\hyperref[fig:conv-se]{4(c)} reports the invariant semantics of the Conv
operator in \F~\hyperref[fig:conv-se]{4(a)}, in the form of a symbolic
constraint.}

\noindent \textbf{General Workflow.}~\mr{Recovering dimensions and parameters
has several tasks. The essential of our solution is to summarize operator
invariant semantics with symbolic execution. We first log execution traces and
use taint analysis to shorten the traces. We then use symbolic execution to
summarize the input-output constraint of each assembly function, infer
dimensions using patterns defined over constraints, and further extract
parameters. We detail each task below.}

\subsubsection{Trace Logging and Taint Analysis}
\label{subsubsec:trace}

Execution trace-based analysis is ideal to analyze DNN executables, because any
non-trivial inputs achieve \textit{full coverage}. \mr{We use Intel
Pin~\cite{pin} to log the execution trace of an operator's assembly function.}
Complex DNN operators like Conv are computation intensive, and a single Conv
execution trace can reach to \textit{hundreds of gigabytes}. Pin takes several
hours to log one trace. Nonetheless, Conv is generally compiled into nested
loops. Hence, analyzing a subtrace containing \textit{one} iteration of the
outermost loop is sufficient (\mr{as long as a complete calculation of an output
element is reflected in this subtrace}).

\noindent \textbf{Taint Analysis.}~\mr{The subtrace can still be up to several
gigabytes in size.} We further use backward taint
analysis~\cite{schwartz2010all,kang2011dta} to rule out instructions that are
\textit{not} involved in computing outputs. We mark this operator's output
elements as taint sources and analyze the trace backward. Our taint propagation
is straightforward to track data
dependency~\cite{schwartz2010all,kang2011dta,wang2017imf}. Trace logging records
each instruction's execution context, including concrete memory address values.
Thus, for each memory access during taint propagation, we compute concrete
addresses to taint/untaint memory cells accordingly.

\subsubsection{Symbolic Execution (SE)}
\label{subsubsec:se}

We launch SE over tainted x86 instructions. \mr{While existing symbolic
execution tools do not support pervasive SSE instructions in DNN executables, we
reimplement a trace-based SE engine that models all SSE floating-point
computations encountered in tainted traces. We ignore irrelevant semantics like
CPU flags.}
As with taint analysis, symbolic pointers are computed using concrete values. For instance, \texttt{movss xmm1, dword ptr [rcx]} will
load floating numbers from memory pointed by \texttt{rcx}. Given that
\texttt{dword} denotes 4 bytes, if \texttt{rcx} is 0x29b8, we create
\texttt{(0x29b8, 4)} as \texttt{xmm1}'s symbolic value while the upper 16-4
bytes are reset to zero. After performing SE on tainted trace, we
get a (simplified) symbolic constraint as in \F~\hyperref[fig:conv-se]{4(c)},
which shows how inputs and parameters in memory (see
\F~\hyperref[fig:conv-se]{4(b)}) are used to computing an output.

\noindent \textbf{Identifying Memory Layouts.}~\mr{To determine if each address
in the symbolic constraint points to inputs or parameters, we form a
once-for-all configuration that records the meaning of each argument (inputs or
parameters) of the corresponding assembly function for different operators (see
Appendix~\ref{sec:prototype}). We can collect and identify inputs and
parameters' memory addresses by querying the configuration.} For the constraint
in \F~\hyperref[fig:conv-se]{4(c)}, we will identify memory addresses and
classify them into weights (marked in \textcolor{pptred}{red}) and inputs
(marked in \textcolor{pptdy}{yellow}).
\mr{Furthermore, by logging and identifying all memory addresses accessed during
an operator's computation, we can cluster all addresses of the same parameter to
scope that parameter's memory region (i.e., the starting address and size).}

\subsubsection{Dimension Recovery}
\label{subsubsec:dimension}

\mr{For reverse engineering, heuristic are hardly
avoidable~\cite{wang2015uroboros, wang2017ramblr}. We now present patterns
defined over the extracted constraints, which enable recovering dimensions and
parameter layouts. Without compromising generality, we mainly introduce patterns
we use to recover Conv operator dimensions and layouts. Other operators in our
dataset can be covered smoothly with simpler patterns, as we stated in
\S~\ref{subsec:design-parameter}.}

\noindent \mr{\textbf{Kernel Size $K$, Input Channel $I_{C}$, Zero Padding
$P$.}}~Consider \F~\hyperref[fig:conv-se]{4(c)}, given the \mr{relative} offsets
of four marked input memory addresses are \texttt{[0, 4, 12, 16]}, we can infer
the kernel shape as $2\times2$ (the continuous sequence has length 2),
indicating that $K=2$. Further, we calculate \#input channels $I_{C} =
\frac{4}{2\times 2} = 1$, as to compute one output element, we recognize four
inputs (which belong to one input channel) in the symbolic constraint.
Also, considering the memory layout in figure b, it should be easy to infer that
``12'' denotes the \mr{first element from the next row (element at row 3, column 2)}.
Hence, we can compute the shape of the input matrix as $\frac{12}{4}=3$ where
$4$ denotes the size of one floating number on 64-bit x86 platforms. The input
shape is therefore $3\times3$. As the network topology has been recovered in
\S~\ref{subsec:design-topology}, we compare the output of the prior operator
with the input of this Conv to decide zero padding $P$. Suppose the output shape
of the prior operator is $1\times1$, $P$ is decided as $\frac{3-2}{2}=1$.

\noindent \textbf{Output Channels $O_{C}$.}~To infer $O_{C}$, we re-run Pin
and log all accessed memory locations when executing Conv. Since we do not need
to log every instruction and its associated context (which involves lots of
string conversions and I/Os), Pin runs much faster than being used to log
execution traces in \S~\ref{subsubsec:trace}. We then re-launch the analysis
detailed in \mr{\textbf{Identifying Memory Layouts}} \mr{to identify all
addresses belonging to weights and then determine the size of the memory region
that stores weights, which implies the size of weights.} Let the memory region
size be $M_w$, $O_{C}$ can be computed as $\frac{M_w}{I_C \times K \times K}$.

\noindent \textbf{Stride $S$.}
Let the input (output) height be $IH_i$ ($OH_i$), we compute
stride $S$ using the following dimension constraint:

\vspace{-10pt}
$$OH_i = [(IH_i + 2P - K)/S] + 1$$
\vspace{-10pt}

The memory region size $M_i$ of Conv inputs can be decided in the same way as
deciding $M_w$. Hence, the input height can be computed as $IH_i=
\sqrt{\frac{M_i}{I_c}}$. Similarly, \mr{we can compute $OH_i$ as
$\sqrt{\frac{M_o}{O_c}}$, where $M_o$ is the memory region size of Conv
outputs.} Stride $S$ is thus computed using the constraint above.

\mr{In sum, \tool extracts a series of facts based on the symbolic constraint
and runtime information of DNN executables, including 1) the memory regions size
of inputs, outputs, and parameters, 2) the relative offsets of the input memory
addresses, 3) the relative offsets of the parameter memory addresses, and 4) the
number of specific arithmetic operations inside a symbolic constraint. Lacking
any of these cannot fully recover a (complex) DNN operator.
Nevertheless, since these facts are consistantly presented in DNN executables
generated by different DL compilers, our dimension recovery techniques are
generic across compilers and optimizations. See Appendix~\ref{sec:recovery} for
handling other operators (in the same procedure) and \S~\ref{subsec:robustness}
for the generalization evaluation.}

\subsubsection{Recover Parameters}
\label{subsubsec:parameter}

Recovering parameters requires to identify their starting addresses and memory
layouts. As clarified in \textbf{\#Output Channels $O_{C}$}, we identify memory
region $M_w$ that stores parameters. 
We then use Pin to dump parameters to disk at runtime.
With recovered dimensions and dumped
parameters in data bytes, we can recover well-formed parameters. Operators may have
multiple parameters, and more than one pointers to distinct parameters may appear
in the assembly function arguments (see Appendix~\ref{sec:prototype} for
function interfaces of operators). Each pointer's parameter is recovered separately. 

\noindent \textbf{Handling Compiler
Optimizations.}~\F~\hyperref[fig:conv-se]{4(b)} depicts a Conv's memory layout.
However, compilers may optimize Conv to reduce runtime cost. Both TVM and Glow
may perform layout alteration optimizations to take advantage of SSE parallelism
by reading 4 (or 8) floating numbers from contiguous memory into one register.
These floating numbers can be computed with one SSE instruction (optimized
memory layout is depicted in \S~\ref{subsec:root-cause}). These
optimizations modify Conv's standard memory layout, impeding parameter recovery.
Similar to dimension inference, we use patterns to identify optimized layouts.
We detail patterns in Appendix~\ref{sec:layout}. In short, \tool\ is nearly
flawless; see discussion in \S~\ref{subsec:root-cause}.
 \begin{figure}[t]
    \centering
    \includegraphics[width=\linewidth]{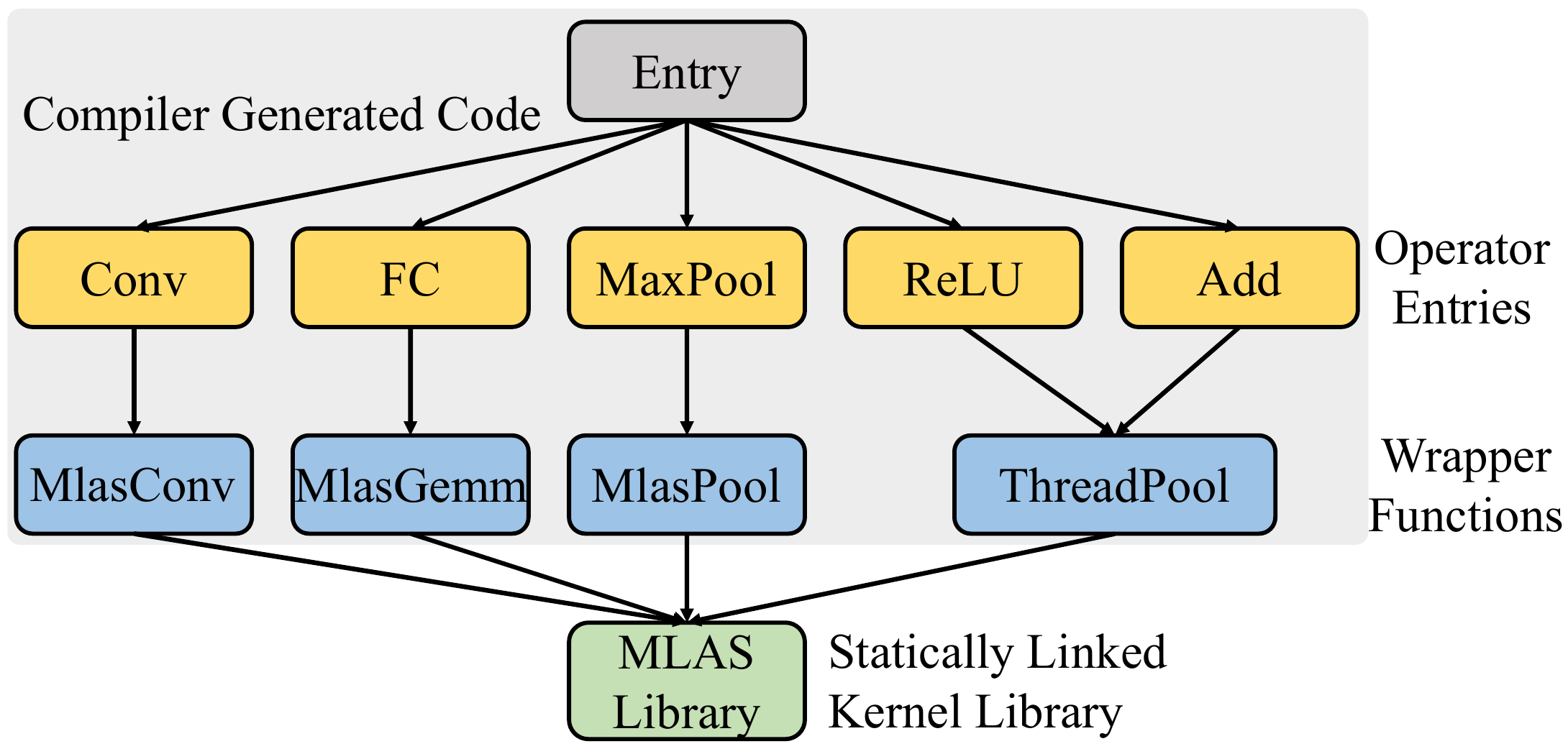}
    \vspace{-20pt}
    \caption{Holistic view of call graph of binary code generated by NNFusion.}
    \label{fig:nnfusion}
  \end{figure}

\subsection{Executables Emitted by NNFusion}
\label{subsec:app-nnfusion}

The procedure described in \S~\ref{sec:design} decompiles self-contained DNN executables --- outputs of the
dominant compilers TVM and Glow. As introduced in \S~\ref{sec:preliminary}, some
DL compilers, including NNFusion~\cite{ma2020rammer} and XLA~\cite{xla},
generate executables statically linked with kernel libraries. It is
\textit{easier} to decompile NNFusion- and XLA-emitted executables since they
contain wrapper functions to invoke target operator implementations in kernel
libraries. We easily determine DNN operator types by matching those wrapper
functions. During runtime, we use Pin to recover network topology and intercept
data sent via wrappers. The intercepted data include dimensions (in integers)
and pointers to parameters. We use obtained dimensions/pointers to load
parameters from the memory. \S~\ref{subsec:eval-nnfusion} assesses \tool\ using
NNFusion-compiled executables.

\mr{\F~\ref{fig:nnfusion} depicts a high-level workflow of DNN executable
compiled by NNFusion, where a set of \textit{distinguishable} wrapper functions
are deployed as trampolines to invoke target DNN operator implementations in
Mlas~\cite{mlas}, high-speed linear algebra library. Since code in kernel
libraries is constant from model to model, we can easily distinguish between the
corresponding wrapper functions by the functions in the libraries being called.
Therefore, it becomes much easier to recover the DNN structure, as we can decide
the DNN operator types via statically pattern matching wrapper functions (e.g.,
\texttt{MlasGemm}). More specifically, we envision that well-developed code
similarity techniques~\cite{ding2019asm2vec,duan2020deepbindiff} or manually
defined patterns could be used to identify kernel library functions smoothly.
While this work primarily focuses on decompiling more challenging code
generation patterns (as elaborated in \S~\ref{sec:preliminary}), given such
patterns are adopted by industrial-strength DNN compilers,
TVM~\cite{chen2018tvm} and Glow~\cite{rotem2018glow}, we only empirically
demonstrate the feasibility of decompiling NNFusion executables in
\S~\ref{subsec:eval-nnfusion}. Executables compiled by XLA can be processed
similarly.}
\section{Implementation}
\label{sec:implementation}

\tool\ is primarily written in Python with about 11K LOC. Our Pin plugins
contain about 3.1K C++ code. The current implementation decompiles 64-bit
executables in the ELF format on x86 platforms, See discussion on cross-platform
support in \S~\ref{sec:discussion}. We use LSTM for DNN operators inference in an ``out-of-the-box'' manner
to deal with distinct optimized low-level code of the same type of operator
resulting from different dimensions. The model is a
one-layer LSTM~\cite{hochreiter1997long} whose hidden dimension is
128. The LSTM is implemented using PyTorch~\cite{paszke2019pytorch}, with CUDA
10.0~\cite{cuda} and cuDNN~\cite{chetlur2014cudnn}. 
\section{Usage \& Error Fixing}
\label{sec:usage}

\fixme{\tool\ offers an end-to-end, automated decompilation. All tasks of
\F~\hyperref[fig:workflow]{3(a)} require no human intervention. However,
decompilation is inherently challenging, and \tool\ may make mistakes. This
section first explains how a user use \tool\ in practice, and then discuss error
fixing.

\noindent \textbf{Usage.}~Given a DNN executable, a user first disassembles it
(e.g., using IDA-Pro) and recovers all assembly functions. The user also need to
provide a format-valid input of this executable for use. Next, as an end-to-end
procedure, \tool\ predicts compilation provenance and each disassembly
function's operator type. \tool\ then launches the network topology recovery
before conducting symbolic execution and recovering dimensions and parameters
for each operator, as explained in \S~\ref{sec:design}. Note that at this step,
\tool\ uses a set of error detection rules (see below) to detect and fix
potential errors. Decompilation process is then re-invoked if errors are fixed.
If the error cannot be resolved, human intervention is required. The user needs
to read and understand the symbolic constraints to fix the error. Human
comprehension at this step is the only uncertain but necessary step of the
decompilation if complex errors occur. Finally, the user can rebuild the model
using the recovered model specification on DL frameworks like PyTorch.}

\mr{\noindent \textbf{Error Fixing.}~To augment \tool's decompilation pipeline,
we provide a set of rules based on the basic knowledge of ML models, whose
violation uncovers decompilation errors. Some rules have error-fixing actions,
but not all errors can be fixed in an automated manner. In that case, the user
can inspect and fix those errors manually. Also, the user may extend the rules
based on their observation and experience of ML models. Currently, we have six
error detection rules, of which Rules 1--4 have follow-up automated fixing
actions, while Rules 5--6 require human intervention for error fixing:

\vspace{-1pt}
\begin{enumerate}[noitemsep,topsep=0pt,leftmargin=6mm]
    \item Dimensions of Conv operators must be integers. Otherwise, \tool reports
    an error and instead uses input/output of predecessor/successor operators to form the
    dimensions; see a relevant case study in Appendix~\ref{sec:err-fix-case}.
    \item Inputs of Add operators must be other operators' outputs.
    If not, \tool infers this operator type as BiasAdd.
    \item The Split operator's output memory region size should be smaller than
    its input memory region size. If not, \tool\ instead infers this operator type as Concatenate.
    \item The symbolic constraint of an operator with ReLU label (e.g., ``Conv+ReLU'')
    must contain a ``max'' operation. If not, \tool\ fixes its inferred operator
    type by removing the ``ReLU'' label (e.g., ``Conv+ReLU'' $\rightarrow$ ``Conv'').
    \item If operator inference model's confidence score is below 80\%, and no error
    is detected by Rules 2-4, \tool throws an error and requires human
    intervention.
    \item An operator's input shape must match its predecessor's output shape.
    Otherwise, \tool throws an error and requires human intervention.
  \end{enumerate}
\vspace{-1pt}

Conceptually, Rule 1 and 6 validate dimension inference, whereas Rules 2--5
validate operator inference results. Rules 2--5 are designed based on
observations of our manual exploration, not failed cases when inferring test
models.}
 
\begin{table}[!htbp]
	\centering
	\scriptsize
  \vspace{-5pt}
  \caption{\mr{Compilers evaluated in our study.}}
	\label{tab:compiler}
	\resizebox{1.00\linewidth}{!}{
		\begin{tabular}{l|c|c|c}
			\hline
      Tool Name & Publication & Developer & Version (git commit) \\
			\hline
			\multirow{3}{*}{TVM~\cite{chen2018tvm}} & \multirow{3}{*}{OSDI '18} & \multirow{3}{*}{Amazon} & v0.7.0 \\
                                   &          &        & v0.8.0 \\
                                   &          &        & v0.9.dev \\
			\hline
      \multirow{3}{*}{Glow~\cite{rotem2018glow}} & \multirow{3}{*}{arXiv} & \multirow{3}{*}{Facebook} & 2020 (07a82bd9fe97dfd)\\
                                   &          &          & 2021 (97835cec670bd2f)\\
                                   &          &          & 2022 (793fec7fb0269db)\\
			\hline
			\multirow{2}{*}{NNFusion~\cite{ma2020rammer}} & \multirow{2}{*}{OSDI '20} & \multirow{2}{*}{Microsoft} & v0.2 \\
                                   &          &           & v0.3 \\
			\hline
		\end{tabular}
	}
  \vspace{-15pt}
\end{table}

\begin{table*}[!htbp]
	\centering
\caption{\mr{Statistics of DNN models and their compiled executables evaluated in our study.}}
	\label{tab:dnn}
	\resizebox{0.77\linewidth}{!}{
		\begin{tabular}{l|c|c|c|c|c|c|c|c}
			\hline
      \multirow{2}{*}{Model} & \multirow{2}{*}{\#Parameters} & \multirow{2}{*}{\#Operators} & \multicolumn{2}{c}{TVM -O0} & \multicolumn{2}{|c}{TVM -O3} & \multicolumn{2}{|c}{Glow -O3} \\\cline{4-9}
      &                    &  & Avg. \#Inst. & Avg. \#Func. & Avg. \#Inst. & Avg. \#Func. & Avg. \#Inst. & Avg. \#Func. \\
			\hline
			Resnet18~\cite{he2016deep}  &  11,703,912 & 69 & 49,762 & 281 & 61,002 & 204 & 11,108 & 39 \\
			VGG16~\cite{simonyan2014very}  & 138,357,544 & 41 & 40,205 & 215 & 41,750 & 185 &  5,729 & 33  \\
			FastText~\cite{bojanowski2017enriching} &   2,500,101 &  3 &  9,867 & 142 &  7,477 & 131 & 405 & 14 \\
Inception~\cite{szegedy2016rethinking}  &  6,998,552 &105 & 121,481 & 615 & 74,992 & 356 & 30,452 & 112 \\
      Shufflenet~\cite{zhang2018shufflenet}  &  2,294,784 &152 & 56,147 & 407 & 34,637 & 228 & 33,537 & 59 \\
      Mobilenet~\cite{howard2017mobilenets}  &  3,487,816 & 89 & 69,903 & 363 & 46,214 & 228 & 37,331 & 52 \\
      Efficientnet~\cite{tan2019efficientnet}  &  12,966,032 & 216 & 89,772 & 546 & 49,285 & 244 & 13,749 & 67 \\
      \hline
		\end{tabular}
	}
\end{table*}

\section{Evaluation}
\label{sec:evaluation}
\fixme{In this section, we evaluate \tool\ by exploring the following four
research questions (RQs) below:}

\mr{\noindent \textbf{RQ1 (Comprehensiveness and Correctness)}: \textit{Is
\tool\ comprehensive and correct to process all operators used in common DL
models compiled with different compilers and optimization options?} 

\noindent \textbf{RQ2 (Robustness)}: \textit{Is \tool robust to survive frequent DL
compiler implementation changes?}

\noindent \textbf{RQ3 (Extensibility)}: \textit{Can \tool be easily extended to
support new operators and models? What efforts are needed?} 

\noindent \textbf{RQ4 (Error Fixing)}: \textit{How does \tool\ handle
decompilation errors?}}

\fixme{We evaluated \tool\ with seven real-world CV models and an NLP
model compiled with eight versions of compilers to provide a comprehensive
evaluation. \tool can produce correct model specifications on 59 of 65 DNN
executables, and experienced users can quickly fix 3 of 6 remaining errors.
Nevertheless, we recognize that some errors cannot be easily fixed by normal
users. In the evaluation, we only use ground truths to verify the correctness of
decompilation results. \tool\ is designed to cope with real-world settings and
does not rely on any ground truth. Setup and results are below.}

\noindent \textbf{Compilers.}~\fixme{\T~\ref{tab:compiler} lists compilers we
used. We select eight versions of three state-of-the-art DL compilers. Glow does
not have a release yet, so we use three versions that are at least six months
apart. Glow and NNFusion only generate fully-optimized executables, but TVM can
be configured to use different optimization levels. Therefore, we use TVM with
no and full optimizations to build two sets of executables, while using Glow and
NNFusion with default settings. All models are compiled into 64-bit x86
executables. \S~\ref{sec:preliminary} and \S~\ref{subsec:app-nnfusion}
describe NNFusion's distinct code generation paradigm. We study decompiling
NNFusion-emitted executables in \S~\ref{subsec:eval-nnfusion}. Other
evaluations in this section use TVM- and Glow-compiled executables.}

\mr{\noindent \textbf{Test Data.}~\T~\ref{tab:dnn} lists all evaluated DNN
models. All these models, except FasstText, are extensively used in CV tasks.
NNFusion can only compile its own shipped VGG11. Our large-scale dataset
includes a total of 675 operators and more than 178 million
parameters. Note that these operators have covered \textit{all} types of DNN
operators used in the CV models in ONNX Zoo. FastText is a common NLP model that
contains Embedding, FC, and Pooling. Embedding is a frequently-used DNN operator
in NLP models that encodes text into embedding vectors. Since Embedding is not
included in the training dataset, we manually label functions in FastText. 

These ONNX files are compiled with TVM and Glow and then disassembled with IDA-Pro.
\T~\ref{tab:dnn} presents assembly code statistics. In general, one (or several
fused) operator corresponds to one TVM/Glow compiled assembly function. Besides,
TVM and Glow will add utility functions (e.g., for memory management).
\T~\ref{tab:dnn} reports average statistics across different versions of DL
compilers.} 

\mr{\noindent \textbf{Training the DNN Operator Identifier.}~To train the DNN
operator identifier, we form a dataset using all \fixme{15} image classification
models (we have excluded image classification models that are in our test
dataset) provided by ONNX Zoo~\cite{onnxzoo}, such as
AlexNet~\cite{krizhevsky2012imagenet} and Inception~\cite{szegedy2015going}.
These models are all commonly-used in daily DL tasks.
To prepare training data, we use TVM and Glow to compile the ONNX files of these
models into executables. DL compilers can be configured to output rich meta
information during compilation, which describes the topology of the compiled
model and dimensions/types information of operators. We take this meta
information as the ground truth.}

\noindent \textbf{Processing Time.}~All experiments run on Intel Xeon CPU
E5-2678 with 256GB RAM and an Nvidia RTX 2080 GPU. Generally, processing
time is not a concern for \tool. Training operator identifier described in
\S~\ref{subsec:design-operator} takes less than one
hour. The total operator identification time is about one second. Recovering DNN network structures
(\S~\ref{subsec:design-topology}) requires only a few seconds since DNN
executables are all lightweight instrumented in this task.
\mr{Taint analysis can take from minutes to hours, depending on model size.
Symbolic execution and parameter extraction usually take several minutes.
Appendix~\ref{sec:process-time} report details of processing time.}

\noindent \textbf{Boosting DNN Attacks.}~\mr{\tool\ extracts high-level model
specifications from executables, allowing attackers to carry out
\textit{white-box} attacks toward the decompiled DNN models. In contrast, when
attackers can only interact with DNN executables, attackers have to launch
\textit{black-box} attacks. With \tool, we demonstrate two attacks: adversarial
example (AE) generation~\cite{Nguyen2015CVPR} and knowledge
stealing~\cite{hinton2015distilling} in a white-box setting. The results suggest
that the white-box attacks enabled by \tool\ are much more powerful than the
black-box settings. \tool\ enables recovering $151.4\times$ more AEs than the
blackbox setting within 20 minutes, and the knowledge stolen from white-box
models are of much higher quality than from the black-box executables; see
details in Appendix~\ref{subsec:eval-attack}.}

\subsection{\textbf{RQ1}: Correctness and Comprehensiveness}
\label{subsec:comprehension}

\mr{This section answers \textbf{RQ1}. Our evaluation dataset contains landmark
CV models and a common NLP model listed in \T~\ref{tab:dnn}. These CV models
contains all kinds of operators used by CV models in ONNX Zoo. We first present
an in-depth evaluation of decompiling CV models, assessing the correctness and
comprehensiveness of \tool's technical pipeline over all included DNN operators.
We then discuss the comprehensiveness over NLP and audio processing models.}

\begin{table}[t]
	\centering
\caption{\fixme{Average accuracy of DNN operator inference}.}
	\label{tab:acc}
	\resizebox{1.0\linewidth}{!}{
		\begin{tabular}{
      @{\hspace{2pt}}c@{\hspace{0pt}}|
      @{\hspace{2pt}}c@{\hspace{1pt}}
      @{\hspace{2pt}}c@{\hspace{1pt}}
      @{\hspace{2pt}}c@{\hspace{1pt}}|
      @{\hspace{2pt}}c@{\hspace{1pt}}
      @{\hspace{2pt}}c@{\hspace{1pt}}
      @{\hspace{2pt}}c@{\hspace{1pt}}|
      @{\hspace{2pt}}c@{\hspace{1pt}}
      @{\hspace{2pt}}c@{\hspace{1pt}}
      @{\hspace{2pt}}c@{\hspace{1pt}}
      }  
			\hline
      \multirow{2}{*}{Model} & \multicolumn{3}{c}{Glow} & \multicolumn{3}{c}{TVM -O0} & \multicolumn{3}{c}{TVM -O3} \\
      \cline{2-10}                
                             & 2020 & 2021 & 2022       & v0.7 & v0.8 & v0.9.dev       & v0.7 & v0.8 & v0.9.dev \\
      \hline
      ResNet18               & 100\% & 100\% & 100\%          & 99.79\% & 99.84\% & 100\%              & 98.15\% & 99.06\% & 99.69\% \\
      VGG16                  & 100\% & 100\% & 100\%          & 99.95\% & 99.79\% & 99.57\%              & 99.75\% & 100\%   & 100\% \\
      Inception              & 100\% & 100\% & 100\%          & 99.98\% & 99.88\% & 99.98\%              & 100\%   & 100\%   & 100\% \\
      ShuffleNet             & 100\% & 100\% & 100\%          & 99.96\% & 99.82\% & 100\%              & 99.62\% & 99.71\% & 99.31\% \\
      MobileNet              & 100\% & 100\% & 100\%          & 99.35\% & 99.46\% & 99.40\%              & 99.80\% & 100\%   & 100\% \\
      EfficientNet           & 100\% & 100\% & 100\%          & 99.65\% & 99.68\% & 99.59\%              & 99.81\% & 99.91\% & 100\% \\
      \hline
   \end{tabular}
}
  \vspace{-5pt}
\end{table}

\subsubsection{Predicting DNN Operator Type}
\label{subsec:eval-operator}

In our test dataset, Glow-compiled executables have \fixme{14} types of DNN
operators and TVM-compiled executables have \fixme{30}. As introduced in
\S~\ref{subsec:design-operator}, our operator identifier outputs a 1D vector of
14 or 30 elements for each assembly function, where a ``1'' in $k$th element
indicates that this function should be labeled to $k$th operator. We allow
multiple ``1'', because operators can be fused into one assembly function. As a
result, DNN operator inference is performed as two-class classification tasks
over 14 or 30 labels. DL compilers provide the ground truth (function labels).
We report the overall accuracy in \T~\ref{tab:acc}, where prediction of an
function is correct, when \mr{the predicted label describes \textit{exactly the
same} operation as the ground truth label}. We interpret the prediction as
highly accurate. Particularly, we achieve 100\% accuracy for all executables
compiled by Glow. We check all errors in TVM and discuss the root causes as
follows:

\noindent \textbf{Data Bias.}~Conv is commonly used with a following ReLU for
feature extraction. Given ``Conv+ReLU'' patterns are frequent in training data,
a ConvAdd operator emitted by TVM -O3 is mislabeled as ConvAddReLU. In contrast,
Dense is often used in the last few layers of DNN without ReLU. Therefore, when
ReLU is fused with DenseAdd under TVM -O3, our model mislabels DenseAddReLU as
DenseAdd. Unbalanced real-world training data causes these mislabels. Errors may
be eliminated by post-checking if symbolic constraints indicating ReLU (i.e.,
containing ``max'') exists.

\noindent \textbf{Operators with Similar Assembly Code.}~In TVM generated code,
BiasAdd can be predicted as Add, and vice versa. As expected, assembly code of
these two operators are similar. \fixme{Our identifier's confidence scores when
labeling such operators with similar assembly code are close to the decision
boundary, i.e., 50\%. However, for other cases, the confidence scores are
\textit{all} significantly higher. We thus use the error detection method
introduced in \S~\ref{sec:usage} to detect such errors in the early stage. Rules
2--4 in \S~\ref{sec:usage} are sufficient to detect all operator labelling
errors. Morevoer, after applying fixing actions associated with Rules 2--4, we
get the correct results over all models, i.e., \textit{all results in Table 3
become 100\%}. Besides, Rule 6 in \S~\ref{sec:usage} will throw a warning and
require human validation when confidence is lower than 80\%.}

\begin{table}[!htbp]
\vspace{-5pt}
  \caption{\mr{Parameter/dimension inference. Each column reports dimension
  inference accuracy/parameter inference accuracy. The complete data is available
  in \T~\ref{tab:quant-full}.}}
	\label{tab:quant}
	\resizebox{1.00\linewidth}{!}{
		\begin{tabular}{
      @{\hspace{2pt}}c@{\hspace{2pt}}|
      @{\hspace{2pt}}c@{\hspace{2pt}}|
      @{\hspace{2pt}}c@{\hspace{2pt}}|
      @{\hspace{2pt}}c@{\hspace{2pt}}
      }
			\hline
      \multirow{2}{*}{Model} & Glow & TVM -O0 & TVM -O3 \\
            & (2020, 2021, 2022) & (v0.7, v0.8, v0.9.dev) & (v0.7, v0.8, v0.9.dev) \\
			\hline
      ResNet18  & 100\%/100\%  & 92.15\%/99.37\% & 100\%/99.37\%  \\
      \hline
    \end{tabular}
  }
  \vspace{-5pt}
\end{table}

\subsubsection{DNN Network Topology Recovery}
\label{subsec:eval-topology}

Recovering network topology (\S~\ref{subsec:design-topology}) is straightforward
and rapid. To validate correctness, we compare the recovered network topology
with the reference DNN's computation graph for executables compiled by TVM -O0.
For all evaluated DNN models, the recovered network structure is \textit{fully
consistent} with the reference. As for executables compiled by TVM -O3 and Glow,
optimizations can change the high-level graph view of models. Thus, it becomes
difficult to compare the recovered topology with reference models. Nevertheless,
we note that all these test cases are shown as flawless in the recompilation
study (\S~\ref{subsec:eval-recompile}). Therefore, the correctness of topology
recovery for optimized cases is validated.

\subsubsection{Parameter and Dimension Recovery}
\label{subsec:eval-parameter}

\mr{\T~\ref{tab:quant}
reports parameter/dimension recovery accuracy. We only list results for ResNet
(as its recovery at this step has defects). Besides ResNet, the accuracies for
all other models are 100\%, and results are consistent across different compiler
versions; see complete data in \T~\ref{tab:quant-full}.} Except for TVM -O0, it
is difficult to compare the recovered dimensions/parameters with the reference
due to compiler optimizations. Hence, \#failures in \T~\ref{tab:quant} equals
\#dimensions or \#parameters that need to be fixed before the recovered models
can be compiled into executables showing \textit{identical} behavior with the
references. Some operator inference failures do not involve
dimensions/parameters, and are thus not reflected in \T~\ref{tab:quant}. 

Overall, \tool\ can determine dimensions of different DNN operators with
negligible errors over all settings. Four failures in ResNet18 (TVM -O0) are due
to a Conv optimization (see \S~\ref{subsec:root-cause}), while all
dimensions of ResNet18 (TVM -O3) are correctly recovered. Besides, despite huge
volume of parameters in each model, the results are promising. \tool\ failed to
recover about 73K parameters of an optimized Conv operator in ResNet18
(TVM -O3) due to its specially-optimized memory layout; see root causes in
\S~\ref{subsec:root-cause}.

\begin{table}[t]
	\centering
\vspace{-10pt}
  \caption{\fixme{Recompilation. ``NA'' means that some errors are not 
  fixed, thus the rebuilt models manifest inconsistent behavior.
  See full results in \T~\ref{tab:recompile-full}.}}
	\label{tab:recompile}
	\resizebox{0.95\linewidth}{!}{
		\begin{tabular}{
      @{\hspace{2pt}}c@{\hspace{2pt}}|
      @{\hspace{2pt}}c@{\hspace{2pt}}|
      @{\hspace{2pt}}c@{\hspace{2pt}}|
      @{\hspace{2pt}}c@{\hspace{2pt}}
      }
			\hline
      \multirow{2}{*}{Model} & Glow & TVM -O0 & TVM -O3 \\
            & (2020, 2021, 2022) & (v0.7, v0.8, v0.9.dev) & (v0.7, v0.8, v0.9.dev) \\
			\hline
      ResNet18     & 100\% & 100\% (with fixing) & NA $\rightarrow$ 100\% \\
\hline
    \end{tabular}
	}
  \vspace{-10pt}
\end{table}

\subsubsection{Recompilation}
\label{subsec:eval-recompile}

Recompilation is an active field in reverse engineering, though recompiling
decompiled C/C++ code is
challenging~\cite{wang2015uroboros,bauman2018superset,wang2017ramblr,williams2020egalito}.
This section demonstrates the feasibility of recompiling decompiled DNN models.
Recompilation requires a fully fledged decompilation, with the end results again
being a functional executable exhibiting identical behavior with the reference.
This demonstrates the feasibility of DNN model reuse, migration, and patching.
To do so, we re-implement DNN models in PyTorch using recovered DNN models,
then export models as ONNX files and compiled into DNN executables using
the same compilation provenance.

It is \textit{not} desirable to directly compare the recovered high-level model
specifications with the reference model's specifications: compilation and
optimization inevitably change DNN model representation (e.g., fusing
operators). Thereby inconsistency of two high-level specifications does not
necessarily indicate a difference in model outputs. Instead, We compare recompiled
and reference executables directly.
Specifically, we compare the predicted labels and confidence scores yielded by
recompiled and reference executables over every input from validation
dataset. Two executables are deemed identical if labels and confidence
scores are \textit{exactly identical} or with only negligible floating-point
precision loss. 
\fixme{For image classification models, we randomly select 100 images from 100
different categories in ImageNet~\cite{imagenet} to form a validation dataset.}
For FastText, we randomly crafted 50 inputs.

\fixme{\T~\ref{tab:recompile} reports the results (only including ResNet18 as
its recompilation has defects). All recompiled models manifest identical
behavior with references over all inputs in the validation dataset except
ResNet. Errors in ResNet (TVM -O0) can be fixed automatically with error fixing
rules (see \S~\ref{subsec:err-fix}), and we mark ``100\% (with fixing)''. \tool\
fails to detect an error in recovering the parameter layout of a Conv in
ResNet18 (TVM -O3); we mark it ``NA''. To verify the correctness of the
remaining recovered operators in this model, we manually fixed this error with
ground truth and re-ran the recompilation study; this model also gets 100\%
correct outputs, marked as ``NA $\rightarrow$ 100\%'' in \T~\ref{tab:recompile}.
In this case, \tool does not produce the correct and directly usable model
specification, and the manual fixing here is merely to prove that all remaining
operators in ResNet18 are correctly decompiled.}

We also measure the size and speed of recompiled and reference executables. We
report that \textit{no} noticeable changes can be observed comparing recompiled
and original executables.

\subsubsection{Decompiling NNFusion Outputs}
\label{subsec:eval-nnfusion}

As clarified in \S~\ref{subsec:app-nnfusion}, decompiling executables emitted by
NNFusion and XLA are much easier, as these executables are linked with kernel
libraries. For completeness, we run an \textit{automated} process to decompile
executables emitted by NNFusion v0.2 and v0.3. NNFusion cannot compile VGG
provided by ONNX Zoo. We thus compile the VGG11 model shipped by NNFusion. This
VGG11 model is in the bytecode format, and we cannot directly compare the
recovered DNN model with the reference. We therefore directly check
recompilation correctness.
When re-compiling the ONNX file of the decompiled model, NNFusion throws
exceptions, which, to our knowledge, seems to be bugs.
To verify the correctness, we instead implement VGG in PyTorch using recovered
VGG descriptions. We follow the same step in \S~\ref{subsec:eval-recompile} to
validate the recovered model. Note that PyTorch and DNN executables may show
negligible deviation between results, which, we believe is from numerical
accuracy instead of errors. We set a threshold to allow 10$^{-4}$ difference in
the outputs of VGG in PyTorch and in executable. All validation inputs are
\textit{passed}. Therefore, we conclude that decompilation is correct.

\begin{figure}[!tph]
  \centering
  \includegraphics[width=\linewidth]{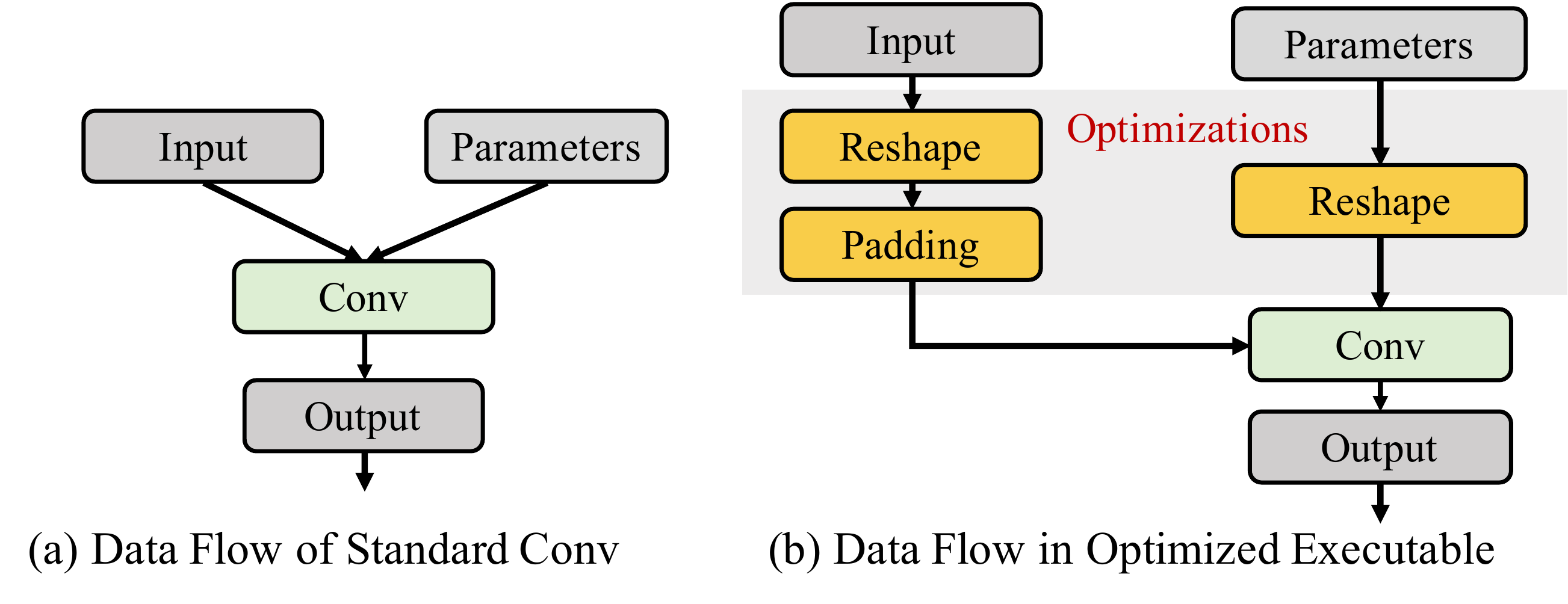}
  \caption{Reshapes inserted before Conv by TVM.}
  \label{fig:case1}
\end{figure}

\begin{figure}[!tbp]
  \centering
  \includegraphics[width=\linewidth]{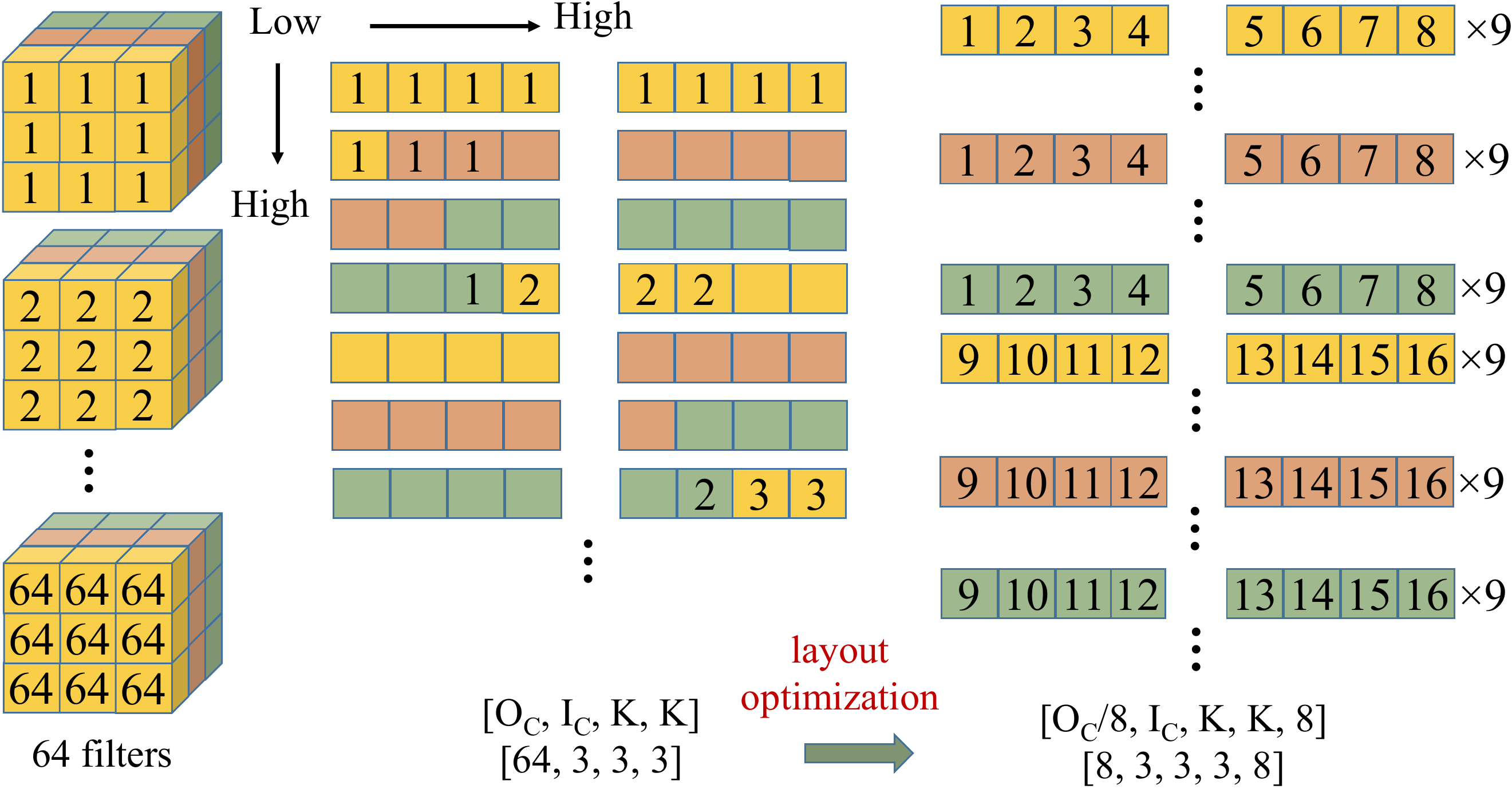}
  \caption{Layout optimization.}
  \label{fig:case2}
\end{figure}

\subsubsection{Root Cause Analysis}
\label{subsec:root-cause}
Besides errors due to mislabeled operators, two failures occurred when inferring
parameters and dimensions. We now discuss the root causes.

\noindent \textbf{Case One.}~All dimension/parameter recovery failures of ResNet
(TVM -O0) are from one Conv operator. The failure is due to a Reshape operator
inserted by TVM before Conv. Recall to infer dimensions of Conv (particularly
the kernel shape $K$), we define patterns over offsets of kernel memory
addresses (\S~\ref{subsubsec:dimension}). However, to speed up program
execution, TVM may add extra code to reshape the input at runtime before
convolutional calculation, as shown in \F~\ref{fig:case1}. In short, after
optimization, the input addresses extracted over symbolic constraints no longer
solely reflect elements in the Conv kernels, thus restraining our patterns.

\noindent \textbf{Case Two.}~Another parameter recovery failure also roots in
Conv compiled with TVM -O3. \F~\ref{fig:case2} presents the standard memory
layout of a Conv which is denoted as $[O_C, I_c, K, K]$. While this layout can
be flawlessly recovered by \tool, as mentioned in \S~\ref{subsubsec:parameter},
DL compilers may alter memory layout of parameters to take advantage of SSE
instructions. As shown in \F~\ref{fig:case2}, the parameter layout can be
converted into an optimized version which is denoted as $[O_C/A, I_c, K, K, A]$.
\tool\ can correctly infer memory layouts for 33 of 34 Conv operators optimized
this way (see our patterns in Appendix~\ref{sec:layout}). Nevertheless, we
still encounter one rare case where weights are not loaded in the order assumed
in our patterns, which is likely due to TVM's auto scheduling. This breaks our
patterns to recover parameters. In contrast, while Glow also extensively uses
this optimization, \tool\ can flawlessly recover parameters from all Conv
operators optimized by Glow.

\subsubsection{Other Models}
\label{subsec:other-recompile}

\begin{table}[t]
	\centering
  \vspace{-10pt}
  \caption{\mr{NLP models compilation results. ``Crash'' means compilers throw
  an exception and terminate. ``Failed'' means executable output is inconsistent
  with the input model. ``Success'' means the model is compiled correctly.}}
	\label{tab:nlp}
	\resizebox{1.0\linewidth}{!}{
		\begin{tabular}{l|c|c|c|c|c}
			\hline
      \multirow{2}{*}{Model} & Glow         & Glow & TVM  & TVM              & NNFusion \\
                             & (2020, 2021) & 2022 & v0.7 & (v0.8, v0.9.dev) & (v0.2, v0.3) \\
			\hline
      Char-RNN~\cite{charrnn} & Success & Success & Failed & Failed & Crash \\
      LSTM~\cite{lstmpytorch} & Crash   & Success & Crash  & Failed & Crash \\
      \hline
   \end{tabular}
}
  \vspace{-5pt}
\end{table}

\fixme{\noindent \textbf{NLP Models.}~We also tried to incorporate NLP models
into our evaluation. However, existing DL compilers still lack complete support
for basic NLP operators, such as RNN and LSTM. \T~\ref{tab:nlp} reports the
results of preliminary investigation. We select two common NLP models,
Char-RNN~\cite{charrnn} and LSTM~\cite{lstmpytorch}, from PyTorch tutorial. Only
the current version of Glow can successfully compile both models. Thus, we
evaluated \tool with Char-RNN compiled with all versions of Glow and LSTM
compiled with Glow 2022, and \tool could smoothly output the correct model
specifications. With manual inspection, we find that typical NLP operators, such
as RNN, GRU, and LSTM, are decomposed into sub-operators during compilation,
including FC operators and Element-wise arithmetic operators. We note that these
decomposed operators are already included in the ONNX Zoo CV
models~\cite{onnxzoo}.}

\mr{\noindent \textbf{Audio Processing Models.}~We expect that \tool\ can also
decompile audio processing models without extension. To clarify, in the era of
deep learning, audios are often converted into 2D representations and then
processed using CV models~\cite{hershey2017cnn}, or directly processed as
sequences using NLP models~\cite{phan2017audio}.}

\begin{tcolorbox}[size=small]
  \mr{\textbf{Answer to RQ1}: \tool\ is correct and comprehensive to cover
  nearly all operators used in common CV models compiled by different compilers
  and optimizations. \tool's applicability for other models is also promising,
  though de facto DL compilers have limited support for them.}
\end{tcolorbox}

\subsection{\textbf{RQ2}: Robustness}
\label{subsec:robustness}

\begin{figure}[!t]
  \centering
  \includegraphics[width=\linewidth]{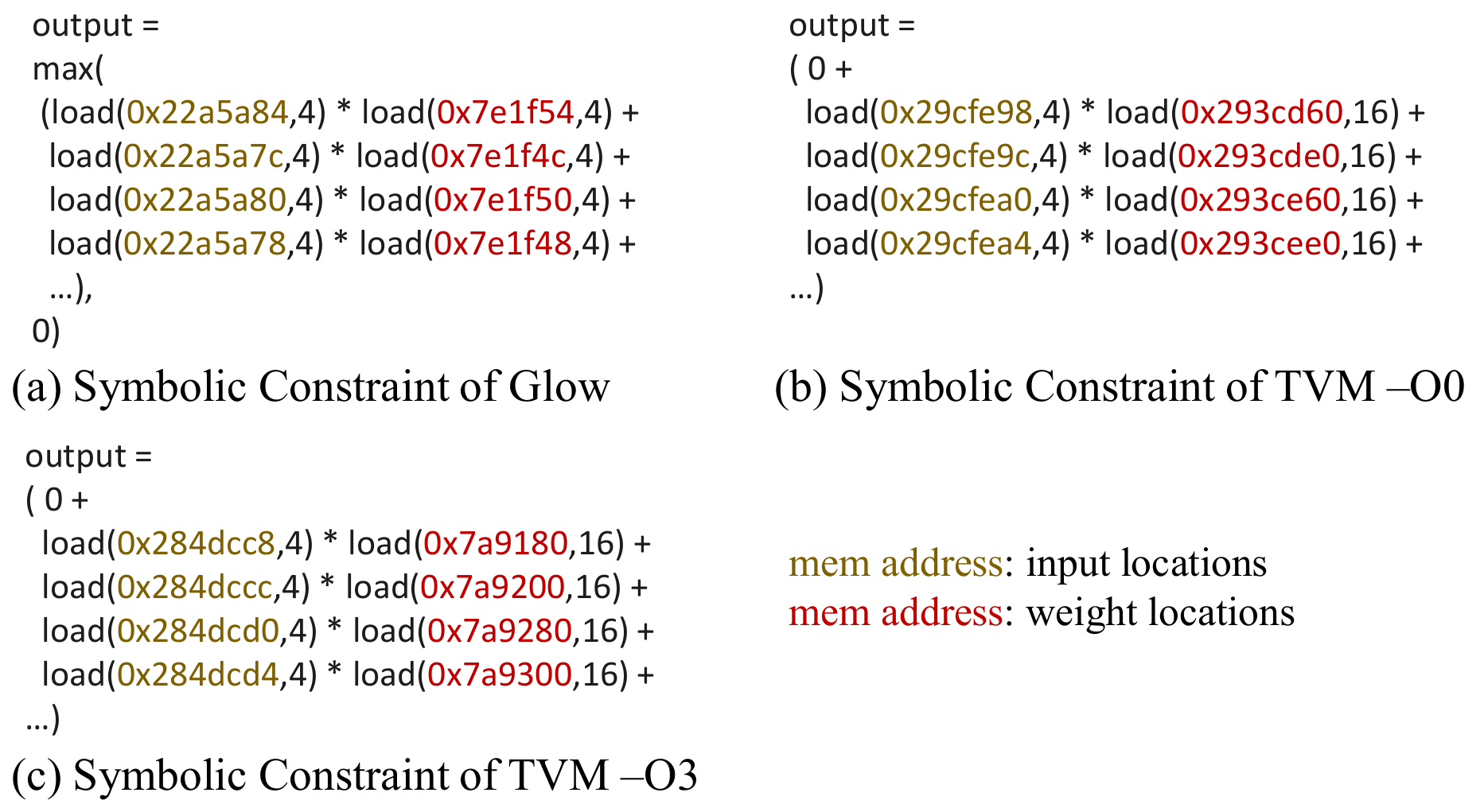}
  \vspace{-20pt}
  \caption{\mr{Mostly consistent symbolic constraints extracted from vastly different binaries.}}
  \vspace{-5pt}
  \label{fig:symexp}
\end{figure}

\mr{\tool\ involves patterns during decompilation. \textbf{RQ2}
arises: \textit{Is this method robust to survive frequent DL compiler
implementation changes?} To answer this question, we evaluated \tool with
prior versions of DL compilers released in the past two years (see \T~\ref{tab:compiler}). 
In short, after testing \tool\ using seven CV models and one NLP model, we
report that \tool\ produces \textit{exactly identical results for different
versions of compilers}. 

We interpret this highly encouraging result from two aspects. First, although
\tool\ leverages patterns to recover dimensions and layouts of parameters, these
patterns are based on semantics constraints, instead of syntax. Since DNN
operators like Conv and ReLU are defined cleanly and rigorously, these
semantics-level information are \textit{consistent} across compiler
implementation changes.
\F~\ref{fig:symexp} illustrates Conv
constraints derived from Glow, TVM -O0, and TVM -O3. Although DL executables are
drastically different, semantic constraints preserve mostly the same pattern.
Notice that \F~\ref{fig:symexp}(a), an extra \texttt{max} exists due to Glow
optimizations. When designing patterns, we deliberately pick components that
\textit{co-exist} across different constraints to recover dimensions and
layouts. Despite complex optimizations imposed by compilers, we find that our
focused components are consistent and robust.}

\begin{table}[t]
	\centering
\caption{\mr{GitHub repo commits investigation results. We report the
  \#Commits in the past two years, \#Commits related to CPU code
  generation, the average LOC per CPU commit, and
  \#CPU commits with substantial changes (over 100 LOC).}}
	\label{tab:commits}
	\resizebox{0.9\linewidth}{!}{
		\begin{tabular}{l|c|c|c|c}
			\hline
      Compiler & \#Commits & \#CPU Commits & Avg. LOC & \#Substantial \\
			\hline
      TVM      & 4,292 & 121 & 17 & 3 \\
      Glow     & 1,435 &  22 & 47 & 3 \\
      NNFusion &   200 &  13 & 16 & 1 \\
			\hline
   \end{tabular}
}
  \vspace{-5pt}
\end{table}

Second, we investigated commits related to CPU instruction generation in
DL compilers' GitHub repos over the past two years (April 2020 to April 2022).
As in \T~\ref{tab:commits}, while these compilers are frequently updated, most
commits aim to increase support for alternative hardware and model formats,
where CPU-related code has changed little. We manually reviewed all
``substantial'' commits, i.e., commits with more than 100 LOC changes, and
confirmed that they do not change optimization strategies or binary code
generation that may affect \tool.
Besides, DL compilers heavily use parallel instruction extensions (e.g., SSE) to
speed up model inference on CPUs. These extensions have been stable and
unchanged over the long term. 
To answer \textbf{RQ2}, we again underline that \tool's essential assumption is
that symbolic constraints extracted from each DNN operator's assembly function
should be invariant across compilers and optimizations. Other features, such as
function signatures, operator fusion, and optimization strategies, are
independent of \tool's core techniques and are also unlikely to be largely
changed in the near future.

\begin{tcolorbox}[size=small]
  \mr{\textbf{Answer to RQ2:} \tool\ is robust enough against changes in current
  and prior versions of DL compilers. We anticipate that compiler changes are
  unlikely to affect the robustness of \tool\ in the near future.}
\end{tcolorbox}


\subsection{RQ3: Extensibility}
\label{subsec:exten}

\begin{table}[t]
	\centering
\vspace{-10pt}
  \caption{Categorize ONNX operators. 
  \texttt{\#with Dims.} denotes \#operators with dimensions to be recovered, 
  \texttt{\#with Opt.} denotes \#operators with compile-time optimizations,
  \texttt{Workload} denotes the amount of work required to cover new operators, 
  and \texttt{\#Covered} denotes \#operators that currently supported.}
	\label{tab:onnxop}
	\resizebox{1.0\linewidth}{!}{
    \begin{threeparttable}
    \begin{tabular}{
      @{\hspace{3pt}}c@{\hspace{3pt}}|
      @{\hspace{3pt}}c@{\hspace{3pt}}|
      @{\hspace{3pt}}c@{\hspace{3pt}}|
      @{\hspace{3pt}}c@{\hspace{3pt}}|
      @{\hspace{3pt}}c@{\hspace{3pt}}|
      @{\hspace{3pt}}c@{\hspace{3pt}}
    }
			\hline
      Category & \#OPs & \#with Dims. & \#with Opt. & Workload & \#Covered\\
			\hline
      Element-wise Op & 56 & 2  &  0 & Low  & 2/2 \\
      Tensor Op       & 32 & 8  &  0 & Low  & 6/8 \\
      Matrix Op       & 2  & 0  &  0 & NA   & NA \\
      Pooling         & 9  & 9  &  0 & Low  & 5/9 \\
      Heavyweight     & 11 & 11 & 11 & High & 7/11 \\
      Normalization   & 6  & 6  &  0 & Low  & 2/6 \\
      Transpose       & 8  & 0  &  0 & NA   & NA \\
      Random          & 6  & 0  &  0 & NA   & NA \\
      Others          & 33 & 0  &  0 & NA   & NA \\
      Total           & 163& 36 & 11 & NA   & 22/36\tnote{*} \\
			\hline
   \end{tabular}
  
    \begin{tablenotes}
      \item[*] 22 of the 36 operators with dimensions that require specific
      patterns to recover are already covered in \tool. 4 of the remaining 8
      operators can be covered with low effort, and the rest 4 (ConvInteger,
      MatMulInteger, QLinearConv, and QLinearMatMul) are rare in common models.
    \end{tablenotes}
  \end{threeparttable}
}
  \vspace{-5pt}
\end{table}

As stated in \S~\ref{subsec:comprehension}, \tool can cover all operators
used in the CV models from ONNX Zoo. This section measures 
\tool's extensibility through the lens of all DNN operators supported by ONNX Zoo
(\textbf{RQ3}). Note that not all operators are for CV models, and not all
operators have been used in DNN models; some of them are rarely used in common
models.
Overall, while most techniques (i.e., operator inference and symbolic execution)
used in \tool are independent of operator types, patterns described in
\S~\ref{subsubsec:dimension} are designed for each complex operator to recover
their parameters/dimensions. Supporting a new operator may need new or
existing patterns. Symbolic constraints are generally human readable, and we 
typically need several hours to design and validate a new pattern for operators 
without complex optimization, like BiasAdd and Pooling. Developing new patterns 
for complex operators like Conv may take days due to complex optimization strategies.

We classified all ONNX operators~\cite{onnxop} to scope \tool's
applicability and the engineering effort required to extend \tool. Consider
\T~\ref{tab:onnxop}, where \texttt{Low} in workload column represents hours of
effort, and \texttt{High} represents several days of work. While ONNX has 163 DNN
operators~\cite{onnxop}, most of them do not have dimensions to be recovered.
Besides, our patterns can be reused with minor modifications to support
currently uncovered tensor operators, pooling operators, and normalization
operators. For heavyweight calculation (Conv, MatMul/FC, GRU, RNN, LSTM, and
their variants), we have already covered 7 out of 11 operators. Note that GRU, RNN, and
LSTM can be covered because they are decomposed into sub-operators, including FC
and element-wise operators, as explained in \S~\ref{subsec:other-recompile}.
Standard models rarely use the remaining operators, including ConvInteger,
MatMulInteger, QLinearConv, and QLinearMatMul. \tool cannot handle these four
operators for the time being, but we expect it will not be challenging to design
patterns for these operators. Essentially, they are variants of Conv and MatMul
operators. We leave support for these operators to our future work.

\begin{tcolorbox}[size=small]
  \textbf{Answer to RQ3}: Users experienced in DL models can spend reasonable
  effort to add support for new operators and models by modifying existing
  patterns in \tool.
\end{tcolorbox}

\subsection{RQ4: Error Fixing}
\label{subsec:err-fix}

This section clarifies how \tool performs error fixing (\textbf{RQ4}). On
one hand, with rules presented in \S~\ref{sec:usage}, \tool can detect and
automatically fix the errors exposed when decompiling the ResNet18 executable
(compiled with TVM -O0). The recovered model specification, after fixing, are
completely correct, as noted in \T~\ref{tab:recompile}. We detail the error
fixing procedure as a case study over the ResNet18 executable in
Appendix~\ref{sec:err-fix-case}.

On the other hand, when errors can not be fixed automatically (ResNet compiled
with TVM -O3), users are required to read the symbolic constraints and fix
errors manually. Generally, fixing an error requires users to be familiar with
both standard operators used in DL models and x86 assembly language.
Nevertheless, even partially recovered models may boost attacks like query-based
model extraction~\cite{tramer2016stealing,papernot2017practical}. 

\begin{tcolorbox}[size=small]
  \textbf{Answer to \textbf{RQ4}:} To cope with decompilation defects,
  \tool\ provides error detection \& automated fixing mechanism, including a
  collection of rules derived from domain-specific knowledge and observations.
\end{tcolorbox}

\section{Discussion}
\label{sec:discussion}

\noindent \textbf{Downstream Applications \& Countermeasures.}~Previous model
extraction attacks rely on repetitive queries or side channels to leak parts of
DNNs. \tool, as a decompiler, reveals a new and practical attack surface to
recover \textit{full} DNNs when DNN executables are accessible.
Appendix~\ref{subsec:eval-attack} will show that \tool\ can boost DNN attacks.
In addition, legacy DNN executables can be inspected, hardened, and migrated to
new platforms. To show the feasibility, we migrated decompiled x86 DNN
executables onto GPUs. This step only requires to use different compiler options
over our recovered DNN models.

DNNs may provide business advantages. Potential security concerns raised by
\tool\ may be mitigated using obfuscation~\cite{larsen2014sok};
particularly, code obfuscation could likely impede DNN operator inference
whereas data obfuscation may likely undermine our patterns over memory layouts.

\noindent \textbf{Cross-Platform.}~As reviewed in \S~\ref{sec:preliminary}, DL
compiler can generate executables on various platforms. The core techniques of
\tool\ are platform independent. We analyze the cross-platform extension of
\tool\ from the following aspects.
First, decompiling DNN executables on devices like hardware accelerators
requires appropriate disassemblers. This demands vendor support and considerable
engineering work. While certain GPU makers like Nvidia provides
disassemblers~\cite{cuda-disassember}, the architecture and ISA of such devices
are only partially or not revealed, preventing migrating \tool\ to these
devices. Intel recently released a dynamic instrumentation tool,
GTPin~\cite{pti-gpu}, but it is immature and limited to Intel processor
graphics. Without vendor support, it is extremely difficult, if not impossible,
to implement \textbf{disassemblers} and \textbf{dynamic instrumentors} on our
own for various devices.

Second, DL compilers produce distinct executables on GPUs and CPUs. For example,
TVM creates a standalone DNN executable on CPU, but a runtime library, including
detailed model information, and an OpenCL/CUDA executable on GPU. Glow has an
immature support for OpenCL using JIT. In short, we see x86 CPU decompilation as
more difficult because inputs are typically \textit{standalone} executables.

\section{Related Work}
\label{sec:related}

\noindent \textbf{Software Reverse Engineering.}~Software decompilation has
achieved primary success. Algorithms are proposed to improve decompiled C/C++
code, including refining type recovery~\cite{lee2011tie,schwartz2018using},
variable recovery~\cite{balakrishnan2010wys,anand2013compiler}, and control
structure recovery~\cite{elWazeer2013scalable,flores2020datalog}. ML accelerates
decompilation~\cite{fu2019coda,hajipour2020ireen,katz2018using,david2020neural}.
Some decompilers are commercially available~\cite{ida,hopper,jeb}. In addition,
recent works have designed decompilers for Ethereum smart
contracts~\cite{zhou2018erays,grech2019gigahorse}. We focus on decompiling DNN
executables by addressing domain challenges to convert DNN executables to
high-level models. We envision that \tool\ will meet demands to comprehend,
exploit, and harden real-world DNN executables.

\noindent \textbf{Model Extraction.}~We have reviewed DL compilation techniques
in \S~\ref{sec:preliminary}. \tool\ enables a novel perspective to extract DNN
models. As introduced in \S~\ref{sec:motivation}, current model extraction works
mostly take ``black-box''
forms~\cite{papernot2017practical,oh2019towards,teitelman2020stealing,orekondy2019knockoff},
where adversaries can assemble a training dataset $(x,y)$ by continuously
feeding inputs $x$ to a target model and collecting its prediction outputs $y$.
The resulting training datasets can be used to train a local model.
Side channels leaked during inference are also used for model extraction,
including timing, cache, and power side
channels~\cite{hua2018reverse,xiang2020open,duddu2018stealing,yan2020cache,hu2020deepsniffer,zhu2021hermes}.
\tool\ is orthogonal to these side channel-based methods. As noted in
\S~\ref{sec:preliminary}, \tool\ can recover full model information whereas they
conduct partial recovery.
We also notice reverse engineering efforts targeting image processing
software~\cite{helium2015pldi,stencil2016pldi,ahmad2019automatically}. These
works use static analysis and heuristics to map (assembly) image processing code
(e.g., blurring) to high-level operators. They analyze image processing software
(e.g., Photoshop), \textit{not} DNN models.
 
\section{Conclusion}
\label{sec:conclusion}

We presented \tool, a decompiler for x86 DNN executables. \tool\ recovers full
DNN models from executables, including operator types, network topology,
dimensions, and parameters. Our evaluation reports promising results by
successfully decompiling and further recompiling executables compiled from
popular DNN models using different DL compilers.
 
\section*{Acknowledgments}
We thank the anonymous reviewers for their valuable comments.
HKUST authors are supported in part by a RGC ECS grant under the contract 26206520.
Lei Ma's research is supported in part by the Canada First Research Excellence
Fund as part of the University of Alberta's Future Energy Systems research
initiative, Canada CIFAR AI Chairs Program, Amii RAP program, the Natural
Sciences and Engineering Research Council of Canada (NSERC No.RGPIN-2021-02549,
No.RGPAS-2021-00034, No.DGECR-2021-00019), as well as JSPS KAKENHI Grant
No.JP20H04168, No.JP21H04877, JST-Mirai Program Grant No.JPMJMI20B8.

\bibliographystyle{plain}

\begin{appendix}

\begin{table}[t]
	\centering
\vspace{-10pt}
  \caption{Average \#instructions in logged and tainted traces.}
	\label{tab:trace-size-full}
	\resizebox{0.95\linewidth}{!}{
		\begin{tabular}{l|c|c|c}
			\hline
      Compiler & Operator & Logged Trace & Tainted Trace \\
			\hline
      \multirow{6}{6em}{Glow} & Conv & 12,494,354    & 104,377 \\
                              & Conv$_{opt}$ & 5,025,031  & 21727 \\
                              & FC   & 212,574    & 8,672 \\
                              & MaxPool   & 1,288,057    & NA \\
                              & AvgPool   & 6,804      & NA \\
                              & LRN       & 1,726,037  & NA \\
			\hline
	  \multirow{7}{6em}{TVM -O0} & Conv & 368,697    & 41,571 \\
                              & FC   & 3,090,359    & 110,643 \\
                              & MaxPool   & 1,365,727    & NA \\
                              & AvgPool   & 48,146      & NA \\
                              & LRN       & 1,537,084   & NA \\
                              & BiasAdd   & 294,656     & NA \\
                              & Embedding & 9,094      & NA \\
			\hline
	  \multirow{7}{6em}{TVM -O3} & Conv & 403,949  & 63,527 \\
                              & FC   & 2,391,832    & 170,336 \\
                              & MaxPool   & 1,374,641    & NA \\
                              & AvgPool   & 193,490      & NA \\
                              & LRN       & 1,577,524    & NA \\
                              & BiasAdd   & 229,042      & NA \\
                              & Embedding & 20,640     & NA \\
			\hline
        \end{tabular}
	}
  \vspace{-5pt}
\end{table}

\begin{table}[!htbp]
\vspace{-5pt}
    \caption{\mr{Parameter/dimension inference. Lines 2--8 report each
    executable's total \#dimensions, correctly-inferred dimensions, and accuracy
    rate for dimension inference. Lines 9--15 report total \#parameters and
    accuracy rate for parameter inference. Different versions of the same compiler
    produce the same results, therefore we merge their columns.}}
    \label{tab:quant-full}
    \resizebox{1.00\linewidth}{!}{
      \begin{tabular}{
        @{\hspace{2pt}}c@{\hspace{2pt}}|
        @{\hspace{2pt}}c@{\hspace{2pt}}|
        @{\hspace{2pt}}c@{\hspace{2pt}}|
        @{\hspace{2pt}}c@{\hspace{2pt}}
        }
        \hline
        \multirow{2}{*}{Model} & Glow & TVM -O0 & TVM -O3 \\
              & (2020, 2021, 2022) & (v0.7, v0.8, v0.9.dev) & (v0.7, v0.8, v0.9.dev) \\
        \hline
        ResNet18  & 65/65/100\%  & 51/47/\textbf{92.15\%} & 78/78/100\%  \\
        VGG16     & 54/54/100\%  & 59/59/100\% & 52/52/100\%  \\
        FastText  & 7/7/100\%    & 7/7/100\%     & 7/7/100\%    \\
        Inception &235/235/100\% & 223/223/100\% &222/222/100\% \\
        ShuffleNet& 82/82/100\%  & 71/71/100\%   & 71/71/100\%  \\
        MobileNet &124/124/100\% & 144/144/100\% &125/125/100\% \\
        EfficientNet & 133/133/100\% & 133/133/100\% & 132/132/100\% \\
        \hline
        ResNet18 & 11,684,712/100\% & 11,703,912/\textbf{99.37\%} & 11,684,712/\textbf{99.37\%}  \\
        VGG16    & 138,357,544/100\% & 138,357,544/100\%  & 138,357,544/100\%  \\
        FastText & 2,500,101/100\% & 2,500,101/100\%    & 2,500,101/100\%  \\
        Inception & 6,998,552/100\% & 6,998,552/100\% & 6,998,552/100\% \\
        ShuffleNet & 2,270,514/100\% & 2,294,784/100\% & 2,270,514/100\% \\
        MobileNet & 3,487,816/100\% & 3,487,816/100\% & 3,487,816/100\% \\
        EfficientNet & 12,950,384/100\% & 12,966,032/100\% & 12,950,384/100\% \\
        \hline
      \end{tabular}
    }
    \vspace{-5pt}
  \end{table}

  \begin{table}[t]
    \centering
\vspace{-10pt}
    \caption{\fixme{Recompilation. ``NA'' means that some errors in DNN models are not 
    fixed, and thus the rebuilt models manifest inconsistent behavior.}}
    \label{tab:recompile-full}
    \resizebox{1.00\linewidth}{!}{
      \begin{tabular}{
        @{\hspace{2pt}}c@{\hspace{2pt}}|
        @{\hspace{2pt}}c@{\hspace{2pt}}|
        @{\hspace{2pt}}c@{\hspace{2pt}}|
        @{\hspace{2pt}}c@{\hspace{2pt}}
        }
        \hline
        \multirow{2}{*}{Model} & Glow & TVM -O0 & TVM -O3 \\
              & (2020, 2021, 2022) & (v0.7, v0.8, v0.9.dev) & (v0.7, v0.8, v0.9.dev) \\
        \hline
        ResNet18     & 100\% & 100\% (with fixing) & NA $\rightarrow$ 100\% \\
        VGG16        & 100\% & 100\%    & 100\% \\
        FastText     & 100\% & 100\%    & 100\% \\
        Inception    & 100\% & 100\%    & 100\% \\
        ShuffleNet   & 100\% & 100\%    & 100\% \\
        MobileNet    & 100\% & 100\%    & 100\% \\
        EfficientNet & 100\% & 100\%    & 100\% \\
        \hline
      \end{tabular}
    }
    \vspace{-10pt}
  \end{table}

\begin{figure}[!t]
  \centering
  \includegraphics[width=\linewidth]{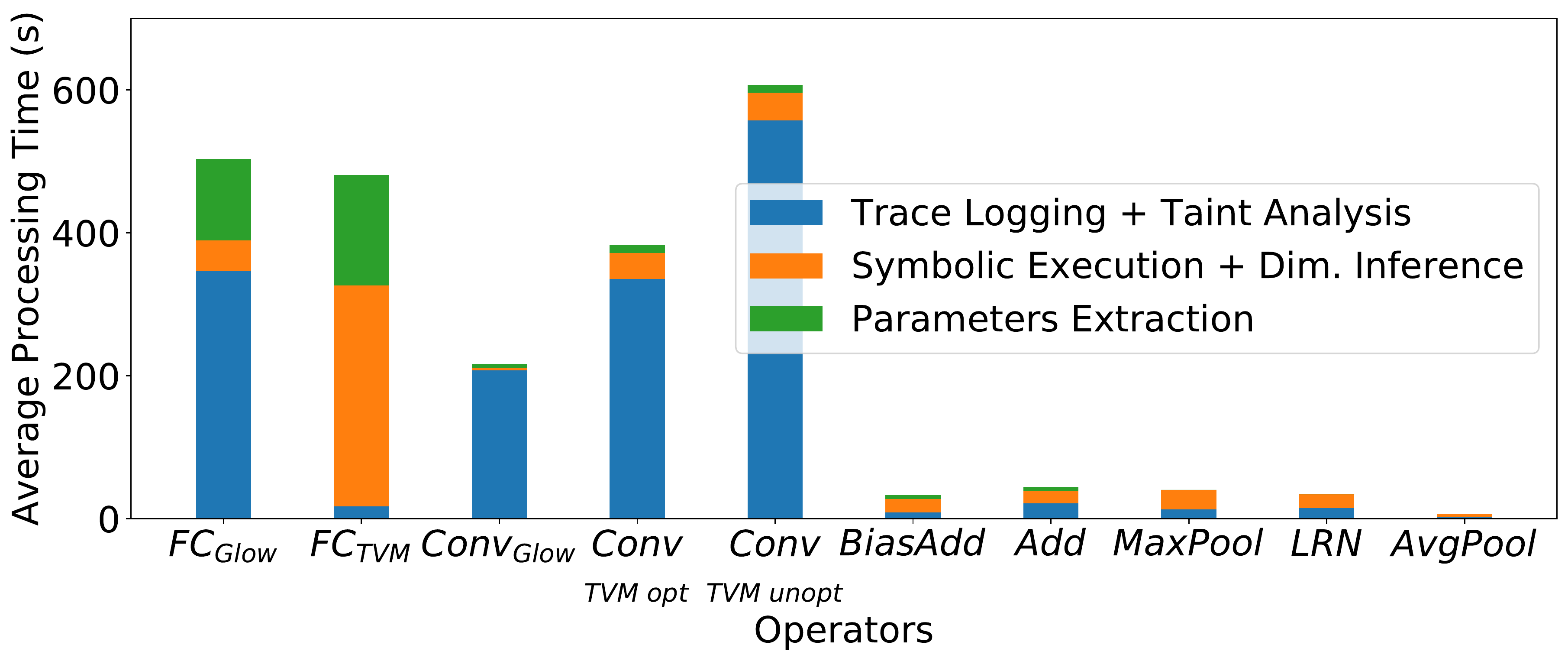}
  \vspace{-20pt}
  \caption{Processing time of parameter/dimension recovery.}
  \vspace{-5pt}
  \label{fig:time}
\end{figure}

\section{Processing Time}
\label{sec:process-time}

Recall \F~\hyperref[fig:workflow]{3(b)} classifies DNN operators into four
types. \F~\ref{fig:time} reports the processing time of recovering parameters
and dimensions over Type \rom{2}, \rom{3}, and \rom{4} operators encountered in
our test cases. One DNN model can contain many instances of an operator, e.g.,
VGG16 has 13 instances of Conv. \F~\ref{fig:time} reports the average cost of
processing one instance.

Trace logging and tainting (\S~\ref{subsubsec:trace}) of Conv and FC operators
can take several minutes; overall, taint analysis itself takes 48.6\% of the
processing time in \textcolor{pyblue}{blue} bars. We further summarize the
length of (tainted) traces in \T~\ref{tab:trace-size-full}.
In evaluation, we configure \tool\ to only conduct taint analysis for Conv and
FC given that they generally emit much lengthy traces than other operators.

\T~\ref{tab:trace-size-full} illustrates that taint analysis is effective to largely
reduce the trace size, which is also partially indicated by comparing
\textcolor{pyblue}{blue} bars and \textcolor{pyorange}{orange} bars in
\F~\ref{fig:time}. Thus, symbolic execution, a more heavyweight analysis, faces
less instructions and is finished rapidly.
Extracting parameters from memory (\textcolor{pygreen}{green} bars) can also
take several minutes. This is because popular CV models usually contain millions
of parameters, and this step involves many I/O operations.
Comparing $FC_{TVM}$ and $FC_{Glow}$, the former takes shorter time to log and
taint traces. Nevertheless, we find that for $FC_{Glow}$, taint analysis taints
only a succinct subtrace. Contrarily, a large portion of instructions are
tainted for the $FC_{TVM}$ case, and therefore, its symbolic execution is still
expensive.

\begin{table}[t]
  \centering
  \scriptsize
  \caption{Configurations required to infer DNN operator parameters from
    corresponding assembly function arguments compiled by TVM.}
  \label{tab:par-tvm}
  \resizebox{0.75\linewidth}{!}{
    \begin{tabular}{l|c}
      \hline
      \textbf{Operator} & \textbf{Configuration} \\
      \hline
      Conv    & in, weights, out \\
      Conv (fused)    & in, weights, biases, out \\
      Dense      & in, weights, out \\
      Dense (fused)      & in, weights, biases, out \\
      BiasAdd     & in, biases, out \\
      AveragePooling & in, out \\
      MaxPooling  & in, out \\
      LRN         & in, out \\
      Sqrt         & in, out \\
      ReLU         & in, out \\
      Divide       & in$_1$, in$_2$, out \\
      Multiply     & in$_1$, in$_2$, out \\
      Negative     & in, out \\
      Flatten      & in, out \\
      Reshape      & in, out \\
      Transpose    & in, out \\
      Transform    & in, out \\
      Concatenate  & in$_1$, in$_2$, out \\
      \hline
    \end{tabular}
  }
\end{table}

\begin{table}[!htbp]
  \centering
  \scriptsize
  \caption{Configurations required to infer DNN operator parameters from
    corresponding assembly function arguments compiled by Glow.}
  \label{tab:par-glow}
  \resizebox{0.90\linewidth}{!}{
    \begin{tabular}{l|c}
      \hline
      \textbf{Operator} & \textbf{Configuration} \\
      \hline
      Conv    & out, in, weights, biases \\
      ConvDKKC8    & out, in, weights, biases \\
      Matmul      & out, in, weights \\
      MatmulAdd   & out, in, weights, biases \\
      MaxPooling     & in, out \\
      AveragePooling & in, out \\
      ReLU       & in/out \\
      BatchAdd  & out, in, biases \\
      Transpose  & in, out, dims  \\
      Reshape    & in, out\\
      Pad        & in, out  \\
      Squeeze  & in, out  \\
      AddReLU   & in$_1$/out, in$_2$  \\
      ExtractTensor  & in, out, offset  \\
      InsertTensor  & in$_1$/out, in$_2$, offset  \\
      LocalResponseNormalization  & out, in \\
      \hline
    \end{tabular}
  }
\end{table}

\section{Assembly Function Prototypes for DNN Operators}
\label{sec:prototype}

As has clarified in the main paper, the calling convention of DNN executables
have eased the task of localizing inputs and parameters for each DNN operator:
we use Intel Pin~\cite{pin} to hook callsites of assembly functions belonging to
the DNN operators, and log the starting addresses of parameters and inputs
during runtime.

With respect to different DNN operators, we pre-define configurations to
describe their assembly function prototypes, particularly the meaning of each
function parameter. \T~\ref{tab:par-tvm} and \T~\ref{tab:par-glow} report the
full configurations in \tool. \T~\ref{tab:par-tvm} describes the assembly
function prototypes in TVM-emitted executables. For instance, the configuration
of Conv indicates that there are three arguments in its compiled assembly
function, where the 1st argument is a pointer toward Conv inputs, the 2nd
argument points to Conv parameters (i.e., weights), and the 3rd argument points
to Conv outputs.
In optimized code (TVM -O3), Conv operators are often fused with following BiasAdd
operators which contribute ``Bias''. As a result, the assembly function of Conv
accepts two pointers toward the global memory regions of weights and biases,
respectively. Similarly, \T~\ref{tab:par-glow} reports the assembly function
prototypes in Glow-emitted executables. Note that some optimized Conv operators are renamed
as ConvDKKC8 by Glow. AddReLU has two inputs, where one input (in$_1$) shares
the memory region with the output.

\section{Parameters \& Dimensions Recovery}
\label{sec:recovery}

\tool\ can recover dimensions and parameters of commonly-used DNN operators. We
\mr{have demonstrated the comprehensiveness by showing that all DNN operators
that exist in all CV models provided by ONNX Zoo~\cite{onnxzoo} can be processed
by \tool.}

\noindent \textbf{General Procedure.}~Due to the limited space, our main paper
only details the recovery of parameters and dimensions for the Conv operator,
which is also the most complex \mr{operator} \tool\ needs to solve. As
introduced in our main paper, we log execution trace using Intel Pin, simplify
the trace with taint analysis, and then conduct symbolic execution to summarize
\mr{invariant semantics in the form of symbolic constraints} over inputs,
parameters, and outputs. We then classify memory
addresses collected over constraints into input addresses and parameter
addresses. This way, we are able to gradually scope the memory layouts of inputs
and parameters, thus inferring dimensions with \textit{patterns}. Parameters
are then dumped from memory to disk \mr{during runtime}, and with the recovered
dimensions, we are able to convert parameters (in data bytes dumped from memory)
into well-formed parameters that can be directly processed by PyTorch.

In addition to solutions of Conv, the rest of this appendix section lists
solutions of all other DNN operators that appear in our dataset. We emphasize
that the \textbf{General Procedure} is indeed applicable to \textbf{all} DNN
operators considered in this work. Given that said, patterns used to infer
dimensions, as expected, need to be defined specifically for different
operators. Also, Conv denotes the most complex DNN operator that requires the
recovery of both parameters and dimensions. Many other operators only contain
dimensions or parameters \mr{and only require simpler but more reliable 
patterns}, as will be clarified below.

\subsection{Fully-Connected (FC) Operator}
\label{subsec:fc}

FC operators are commonly-used to serve the last few operators of many popular
DNN models. In general, neurons in a FC operator are fully connected to all
activations in its previous operator. The activations of neurons in a FC
operator can be typically computed with a \textit{matrix multiplication}
followed by a bias \mr{addition}. Let the input $I$ of a FC operator be a matrix of size $(1,
M)$, and the output $O$ be a matrix of size $(N, 1)$. Hence, parameters $P$ of a
FC operator can be represented as a $(M, N)$ matrix, and Bias $B$ is a matrix of
$(N, 1)$. The output can be calculated as $O=I \times P + Bias$.

\noindent \textbf{Inferring Dimensions $M$ and $N$.}~For dimensions, we need to
first recover the input memory size $M$ and the output memory size $N$. Given
matrix multiplication $O=I \times P$ is performed in a FC operator, it is easy
to see that each element $o$ in the output matrix of size $(N, 1)$ is calculated
using $M$ elements in the input and $M$ elements from the parameters $P$.
Therefore, once we have performed symbolic execution to summarize the symbolic 
constraint over inputs, parameters and one output $o$, it is easy to infer $M$, by 
directly counting the number of \mr{multiplication} that are found in the constraint. 
To infer $N$, we re-run Pin to log all memory writes, scope the size
of the output region $M_o$, and divide $M_o$ by the size of an output element
$o$. The size of $o$ is 4 bytes, denoting a 32-bit floating point number, on
64-bit x86 platforms.

\subsection{Pooling Operator}
\label{subsec:pooling}

A pooling operator downsamples the output of its prior operator. It is commonly
used in dimension reduction to reduce the number of operations required for the
following operators, but still retaining sufficient information from its
previous operators. Despite that there are several different pooling schemes,
e.g., max pooling (MaxPool), average pooling (AveragePool), and min pooling
(MinPool), we infer the dimensions of pooling operators in a unified manner.

\noindent \textbf{Patternss on Dimension Inference.}~In general, for a pooling
operator, we need to recover its stride $S$ and the kernel size $K$. Similar
with computations launched in a Conv operator, the pooling operator will
iteratively extract a sub-matrix, namely \textit{kernel}, for pooling. We
clarify that kernel size $K$ is inferred in exactly the same way as how kernel
size is inferred for Conv. To recover stride $S$, we first launch symbolic
execution and generate the symbolic constraints for two consecutive output
elements, namely $A$ and $B$. The offsets of input memory addresses shall
indicate stride $S$. Specifically, we calculate $S$ through the following
constraint:

$$S= \frac{addr_1 - addr_2}{size}$$

\noindent where $addr_1$ denotes the smallest input address found in constraint
$A$ and $addr_2$ denotes the smallest input address found in constraint $B$.
$size$ denotes the size of one input element, which is 4 bytes in our research
context.

\subsection{Arithmetic Operators}
\label{subsec:add}

In general, arithmetic operators in DNN models can be divided into two
categories: \textit{parameterized} operators and \textit{parameter-free}
operators. For parameter-free operators, such as Add, Sub, Div, Sqrt, which
perform element-wise arithmetic operations over one (for Sqrt) or two inputs.
Since the operations are element-wise, there is no dimension to be recovered.

For arithmetic operators with parameters, the memory starting addresses of
parameters can be obtained from the assembly function inputs (see
\T~\ref{tab:par-tvm} and \T~\ref{tab:par-glow}). Furthermore, we need to recover
the size of parameters in order to dump parameters from memory. For example, DL
compilers will use BiasAdd (or BatchedAdd) to implement the ``add bias''
operation in a Conv operator. To extract biases, we first need to know its size.
However, the size of parameters cannot be inferred from the symbolic
constraints, because the operation is element-wise, thus only one input element
and one parameter will be involved in each symbolic constraint to compute an
output element. Recall when analyzing Conv and FC, we use Pin to record all
memory accesses toward parameters to gradually scope the memory region size of
parameters. Similarly, here we conduct the same process to obtain the memory
region size of biases.

\subsection{Embedding Operator}
\label{subsec:design-nlp}

Despite the popularity of DNN models in CV applications, natural language
processing (NLP) models are also extensively adopted in real-life scenarios.
Despite all common DNN operators that can be handled by \tool, Embedding is
unique to NLP models. Embedding is frequently used to preprocess input text data
by encoding text into embedding vectors. Embedding is essentially implemented as
a hash table that maps semantically similar inputs to similar vectors in the
latent embedding space. To recover an embedding operator, we need to infer the
embedding dimension $D$ (the size of the embedding vector), and the number of
embeddings $N$ (the size of the vocabulary or embedding table). For
Glow-compiled executable, given that \texttt{memcpy} functions are used to
implement Embedding, we can directly hook these \texttt{memcpy} functions and
log its copied memory size, which equals to $D$. For TVM-compiled executable, to
infer $D$, we use Pin to log all memory addresses that are accessed for memory
read. The size of the longest and continuous memory chunk equals to $D$. To
obtain $N$, we record the entire memory region of the embedding table used by
Embedding, the size of which is denoted by $M$. Then, we compute $N$ as
$\frac{M}{D}$.

\subsection{Miscellaneous}
\label{subsec:misc}

\noindent \textbf{Activation Functions.}~Typical DNN models include a large
volume of non-linear activation functions like Sigmoid, Tanh, Softmax, and ReLU.
To ease the presentation, we divide these activation functions into two
categories: \textit{element-wise} activations and \textit{tensor-wise}
activations. As aforementioned, element-wise activations like ReLU, Mish, and
Tanh, will apply activation function one element each time. For cases of this
category, there are no dimensions to be recovered. On the other hand,
tensor-wise activations will apply activation functions to an $n$-dimensional
input tensor, e.g., Softmax, Softmin, and LogSoftmax. For these cases,
recovering the dimension $N$ of the input tensor is straightforward, given that
for each symbolic constraint over inputs and one output element, the number of
involved input elements equal to $N$.

\noindent \textbf{Local Response Normalization (LRN).}~LRN usually follows a
Conv operator, which applies normalization across channels. More specifically,
the dimensions of output is the same as the dimension of input, and each output
element is calculated using input elements from $n$ neighboring channels. While
there is no parameter involved, we need to recover the number of neighboring
channels $n$, as a configuration of LRN. As expected, given a LRN operator whose
number of neighboring channels is $n$, there will be $n$ input elements
involved in the symbolic constraint to compute one output element. Thus, we can
infer $n$ by counting the number of input addresses found in the symbolic
constraint.

\noindent \textbf{Transpose, ExpandDims, Flatten, etc.}~\mr{These operators do
not perform arithmetic operations. Instead, they change the dimensions, alter,
or reshape the layout of inputs. We observed that such operators do not exist in
the original model definition but are extra added by DL compilers to apply
optimization strategies involving memory layout changes, as discussed in
\S~\ref{subsec:root-cause}. Hence, while no parameters are involved in the
computations, and such operators do not even exist in the original model, we
ignore these operators during dimension recovery and model rebuilding. }

\noindent \textbf{Batch Normalization (BN).}~BN normalizes (e.g. scaling and
shifting) operator inputs to enhance model learning quality and robustness. BN
operators typically ship with parameters that were updated during training. The
input and output dimensions are the same.
It is worth noting that BN operators do \textit{not} need to be handled
specifically if the DL compiler optimizations are enabled: according to our
observation, BN operators are typically fused with its prior operators (i.e.,
Conv) in the compiled executable, which changes the values of parameters in
prior operators, but does not impede our solutions.
In case that optimizations are disabled, BN operators are formed by simple
arithmetic and utility operators. For example, when compiling with TVM -O0, the
BN operator will be compiled as a sequence of basic arithmetic operations: [Add,
  Sqrt, Divide, Multiply, ExpandDims, Multiply, Negative, Multiply, Add,
  ExpandDims, Add.] \mr{ExpandDims can be ignored, as discussed above}, and all
of these arithmetic operators can be smoothly decompiled following the solution
detailed in \S~\ref{subsec:add}.

\noindent \textbf{InsertTensor, ExtractTensor, Concatenate, etc.}~\mr{Glow uses
InsertTensor and ExtractTensor to do tensor level operations. For example, Glow
may use InsertTensor to implement padding operations by inserting the input
tensor into another wider tensor filled with zeros. Moreover, TVM uses
Concatenate for similar operations. For these operators, we need to recover the
attributes employed to manipulate tensors, such as offsets. According to our
observation, such attributes are also passed to operators via arguments of
assembly functions. Thus, we are able to get the attributes via instrumentation
without using patterns.}

\begin{figure}[t]
  \centering
  \includegraphics[width=1.0\linewidth]{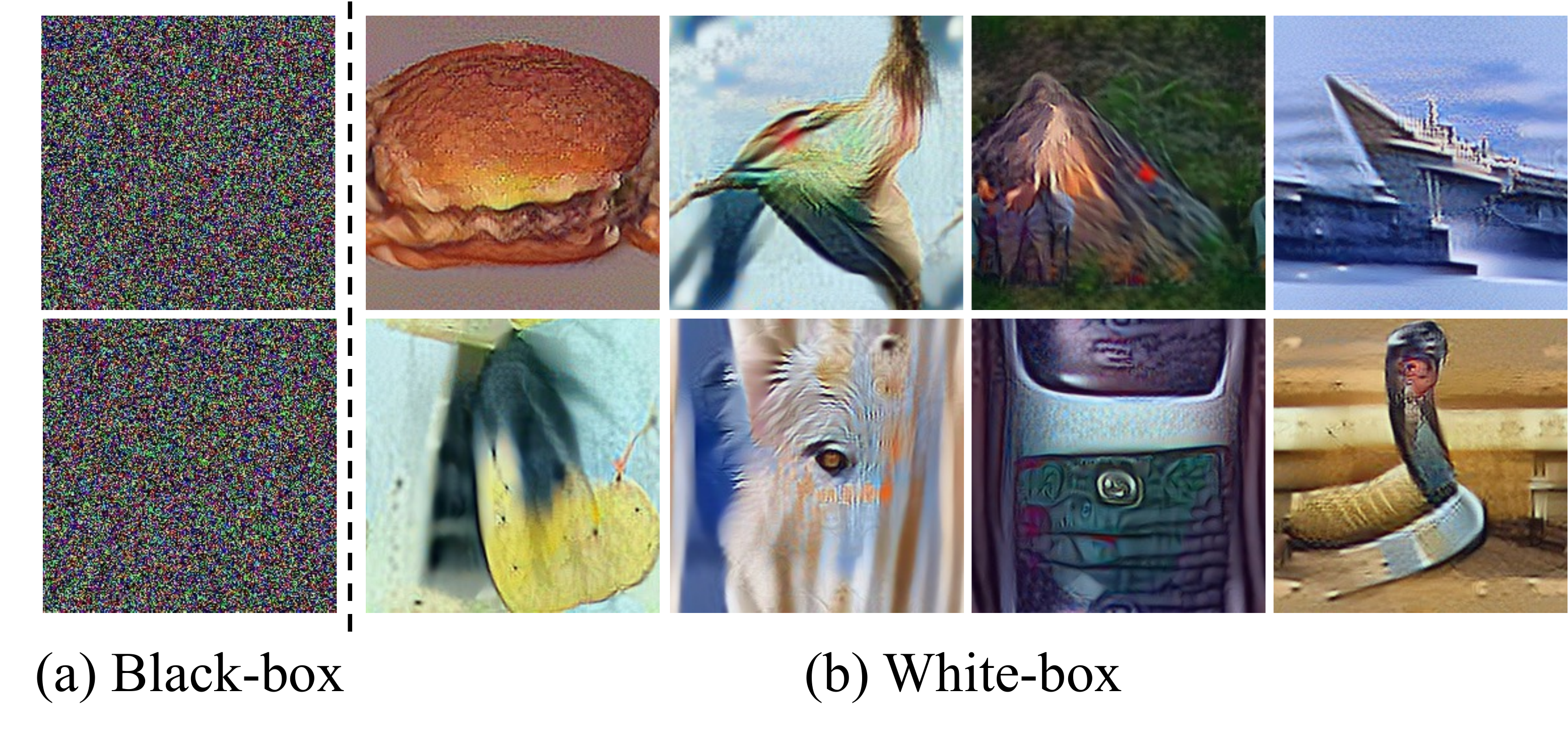}
  \caption{Images extracted from the ResNet18 executable (``black-box'') and
    from the decompiled ResNet18 model running on PyTorch (``White-box'').}
  \label{fig:distill}
\end{figure}

\section{Boosting DNN Attacks}
\label{subsec:eval-attack}

Many studies have been conducted on black-box attacks against DNN
models~\cite{suya2020hybrid,papernot2017practical,yuan2021consistency,yang2020learning,huang2019black}.
However, these methods are usually time and resource intensive. While white-box
attacks are more efficient, attackers may not have access to the white-box
models~\cite{zhang2020two,leino2020stolen}. In our scenario, we assume that
attackers with only DNN executables on hand as a \textit{black-box} setting. In
contrast, by decompiling DNN executables into high-level DNN specifications,
\tool\ enables a \textit{white-box} attack where attackers can inspect the
internals of decompiled DNN models running on DL frameworks like TensorFlow or
PyTorch. At this step, we present two case studies in which we 1) launch
black-box attacks toward ResNet18 executable compiled by TVM -O3, 2) use
\tool\ to decompile ResNet18 executable, and 3) launch white-box attacks toward
the recovered ResNet18 running on PyTorch. We compare the attack effectiveness
to illustrate the improvement of white-box settings over the black-box settings.
We give the major setups and results below; for attack details, see
Appendix~\ref{sec:append-attack}.

\noindent \textbf{Adversarial Example Generation.}~Adversarial examples (AEs)
are crafted inputs with the purpose of confusing a
DNN~\cite{Nguyen2015CVPR,moosavi2017universal}. Typical black-box AE generation
rely on a large number of time-consuming
queries~\cite{papernot2017practical,liu2016delving}. To enhance AE generation on
DNN executables, we first decompile the DNN executable of ResNet18 compiled with
TVM -O3, and deploy the decompiled ResNet18 on PyTorch. We then employ white-box
attack using the gradients acquired from the decompiled ResNet18 model. We
employ the projected gradient descent (PGD) attack~\cite{madry2018towards} for
the white-box scenario and the state-of-the-art (SOTA) AE generation method for
the black-box scenario~\cite{guo2019simple}. Within 20 minutes, we can
successfully generate 34,780 AEs using the white-box attack, compared to only
228 AEs using the black-box attack. The result is intuitive: white-box AE
generation enabled by \tool can largely outperform the black-box attack.

\noindent \textbf{Knowledge Stealing.}~Knowledge stealing, also known as
knowledge transfer~\cite{hinton2015distilling,yin2020dreaming}, is the process
by which knowledge is extracted from a victim DNN model.
DeepInversion~\cite{yin2020dreaming}, the SOTA ``data-free'' knowledge stealing
technique, requires no training data but only the victim model $m$.
DeepInversion synthesizes images from $m$; it is shown that the synthesized
images can be used to train another model with competitive
accuracy~\cite{yin2020dreaming}. It is easy to see that DeepInversion can steal
intellectual property subsumed in well-trained DNN models in the form of
synthesized images. Nevertheless, a DNN executable, denoting a black-box
setting, can hardly enable knowledge stealing since it does not provide
gradients. To enable knowledge stealing, we first use \tool\ to decompile the
ResNet18 executable. We then launch the knowledge stealing process toward the
decompiled ResNet18 (running on PyTorch). We use DeepInversion on both the
ResNet18 executable and the decompiled model. For the black-box setting, we use
a heuristic-based search strategy to replace standard gradient-based methods
(details in Appendix~\ref{sec:append-attack}). As illustrated in
\F~\ref{fig:distill}, DeepInversion is capable of smoothly generating quality
images from the decompiled ResNet18, indicating the success of knowledge
stealing attack. In contrast, in the black-box scenario, only random noise is
generated.

\section{Error Fixing Case Study}
\label{sec:err-fix-case}

\begin{figure}[t]
  \centering
  \includegraphics[width=\linewidth]{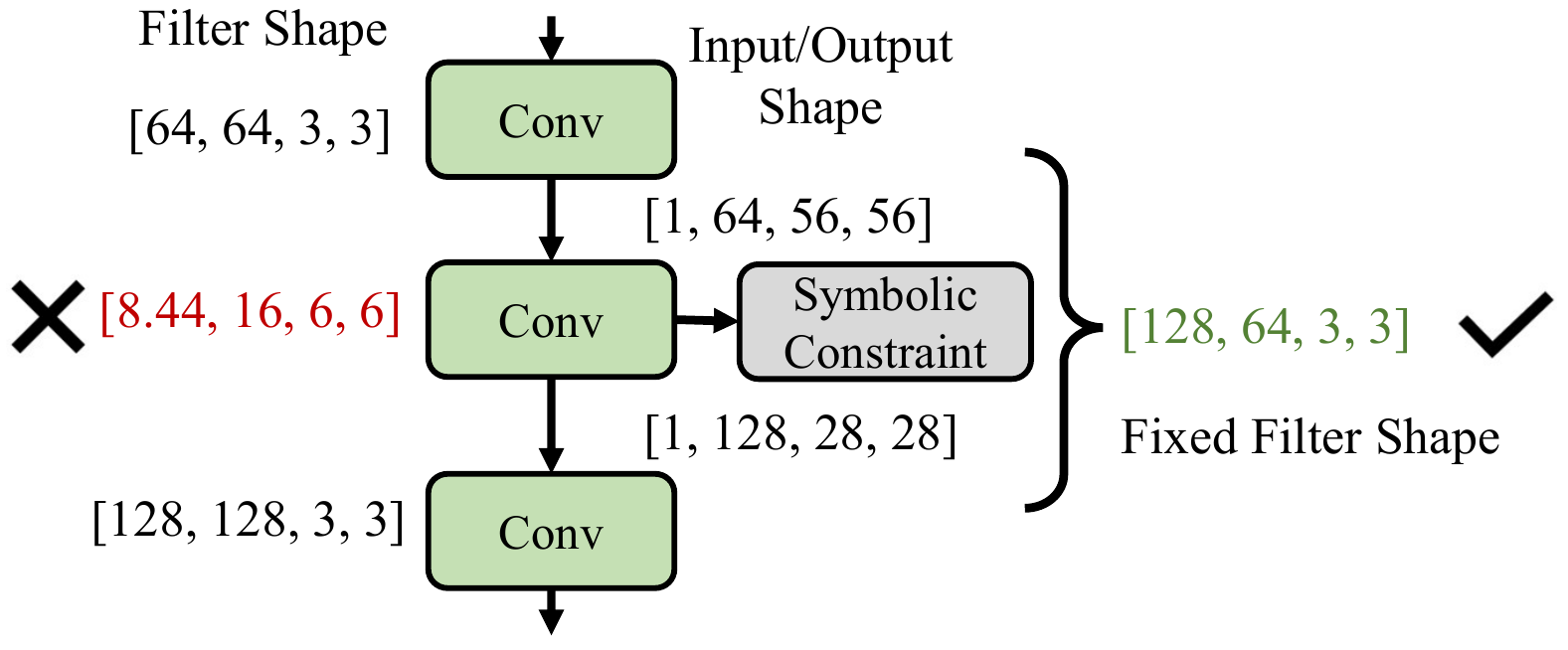}
  \caption{Error Fixing Case Study.}
  \label{fig:err-fix}
\end{figure}

\mr{We introduced the error detection rules we used to automatically detect and fix
errors in \tool's output in \S~\ref{sec:usage}. In this section, we provide a
case study to show why some errors are fixable. \F~\ref{fig:err-fix} shows that
\tool recovers the filter shape of a Conv as $[8.44, 16, 6, 6]$ wrongly, because
of the reshape optimization described in \textbf{Case One} in
\S~\ref{subsec:root-cause}. Such an abnormal shape will be detected as an
error by Rule 1 proposed in \S~\ref{sec:usage}.

To fix this error, Rule 1 leverages input/output shapes from
successor/predecessor operators to determine the input and output shape of the
current Conv operator. As illustrated in \F~\ref{fig:err-fix}, combining the
input shape $[1, 64, 56, 56]$, output shape $[1, 128, 28, 28]$, and symbolic
constraint, we are able to infer the input channel $I_c$ as $64$, and output
channel $O_c$ as $128$. Assume the number of multiplications in constraint is
$n$ ($576$ in this case study), we are also able to infer the kernel size $K$ as
$\sqrt{\frac{n}{I_c}}$, i.e., $3$. Thus, the dimension of the Conv is
automatically fixed as $[128, 64, 3, 3]$.}

\begin{figure*}[!tbp]
  \centering
  \includegraphics[width=0.8\linewidth]{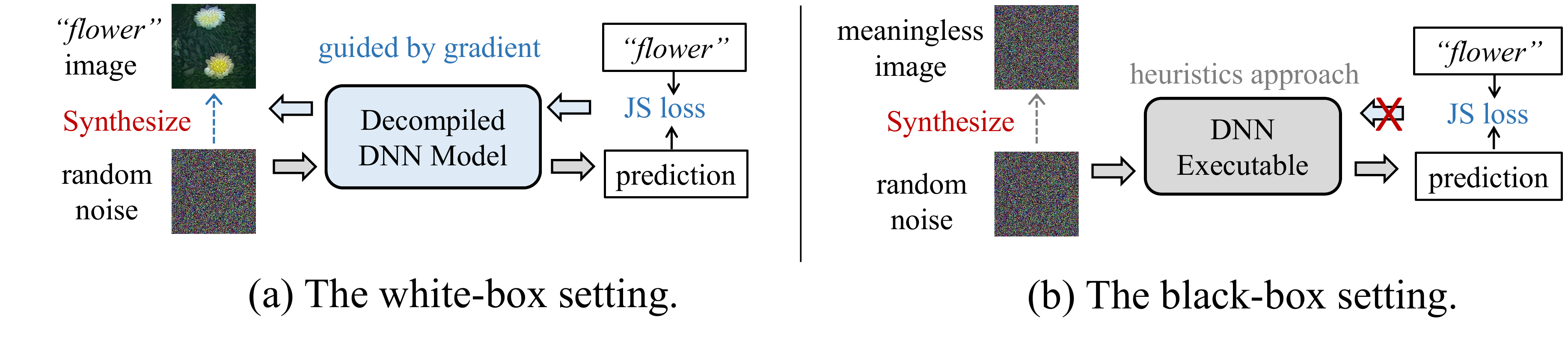}
  \caption{Overflow of data free knowledge distillation.}
  \label{fig:deepinversion}
\end{figure*}

\section{Recovering Parameters From Optimized Memory Layouts}
\label{sec:layout}

Our main paper has clarified that parameters of Conv can be dumped from memory
with its starting address obtained from assembly function inputs together with
the inferred dimensions. However, to take full advantage of SSE parallelism on
x86 platforms, DL compilers may perform layout alteration optimization to change
the standard memory layout of parameters. 
Overall, we find that Glow may change the memory layout from $[O_C, I_C, W, C]$
into $[O_C/A, I_C, K, K, A]$ (as shown in \F~\ref{fig:case2}). Similarly, TVM
may change the layout into a $6$-dimensional tensor $[O_C/B, I_C/A, K, K, A,
  B]$.

Hence, inferring $A$ and $B$ becomes a pre-requisite to comprehend the optimized
memory layouts. Given a summarized symbolic constraint over inputs, parameters,
and one output element, we extract the memory offsets (w.r.t.~the smallest
memory address on the constraint) of all memory addresses belonging to weights,
which will derive the following expression:

\begin{align*}
  [\\
    Y_0[&x_{0,0},\  x_{0,1},\  ...,\  x_{0,(W-1)}, \\
      &x_{1,0},\  x_{1,1},\  ...,\  x_{1,(W-1)}, \\
      &,\  ..., \\
      &x_{(A-1),0},\  x_{(A-1),1},\  ...,\  x_{(A-1),(W-1)} ],\\
    Y_1[& ...,\  ...,\  ...], \\
    &, ..., \\
    Y_{C \times H/A}[&x_{0,0},\  x_{0,1},\  ...,\  x_{0,(W-1)}, \\
      &x_{1,0},\  x_{1,1},\  ...,\  x_{1,(W-1)}, \\
      &,\  ..., \\
      &x_{(A-1),0},\  x_{(A-1),1},\  ...,\  x_{(A-1),(W-1)}]\\
  ]\\
\end{align*}

\noindent where each $x_{i,j}$ is a memory offset of one weight element (i.e.,
the memory address of one small box in the optimized memory layout in
\F~\ref{fig:case2}). Here $Y$ is a notation to ease the representation,
otherwise $x$ will be defined using 3 dimensions.

Furthermore, the optimized memory layout in \F~\ref{fig:case2} can indeed derive
the following constraint

$$Y_k[x_{i,j}]=(i-1) \times B + j \times A \times B + k \times W \times A \times B$$

Hence, $A$ and $B$ can be inferred, by matching the above constraint with two
elements in the extracted memory offsets, e.g., $Y_0[x_{0,1}]$ and
$Y_0[x_{0,3}]$.

For example, let the offset list of weight memory addresses in a symbolic
constraint be
$$[0, 1024, 2048, 32, 1056, 2080, 64, 1088, 2112,...]$$, and the filter shape
$[O_C, I_C, K, K]$ recovered in advance be $[256,128,3,3]$. From the 2nd offset
in the list, we can infer that $A \times B=1024$. Similarly, from the 4th
offset, we can infer that $B=32$. This way, $A$ and $B$ are recovered, and the
layout configuration becomes $[O_C/32, I_C/32, K, K, 32, 32] = [256/32, 128/32,
  3, 3, 32, 32]$. Knowing the layout can enable smoothly loading all parameters
out from memory and re-forming the original filter weights in Conv.

\section{Setup Details of Attacks Boosted by \tool}
\label{sec:append-attack}

For the case study presented in Appendix~\ref{subsec:eval-attack}, the target of
attacks is decided as the DNN executable of ResNet18 downloaded from ONNX
Zoo~\cite{onnxzoo} and compiled by TVM with full optimizations. The attacks are
launched using a Nvidia RTX 2080 GPU.

\noindent \textbf{Adversarial Examples Generation (AE).}~We launch the
adversarial attack with seed images from ImageNet. We reuse the publicly
available implementation~\cite{harry24k} of Projected Gradient Descent
(PGD)~\cite{madry2018towards} over the decompiled ResNet18 model specifications.
This denotes the \textit{white-box} setting where we can fully access model
gradients. Launching a white-box AE generation is standard and straightforward.
The maximum perturbation is set to 0.3, the step size is set to 2/255 (where 255
is the maximal value of a pixel), and the total number of steps is set to 40. As
reported in Appendix~\ref{subsec:eval-attack}, we successfully generate 34,780 AEs
during a 20-minute experiment.

As for the black-box setting where attackers only have the executable file of
ResNet18, we utilize the official release~\cite{cg563} of the state-of-the-art
(SOTA) black-box AE generation attack~\cite{guo2019simple}. We set the batch
size as 50, the maximum perturbation as 0.3, a dimensionality of 2D frequency
space as 14, and a sampling rate of 1,000 for the image samples. As noted in
Appendix~\ref{subsec:eval-attack}, during a 20-minute experiment, we only successfully
generate 228 AEs.

\noindent \textbf{Knowledge Stealing.}~Knowledge stealing, also known as (data
free) knowledge transfer, is a technique to extract knowledge (in terms of
training data) in well-trained victim DNN to train a new ``student'' model. We
employ the official PyTorch implementation~\cite{nvlabs} of
DeepInversion~\cite{yin2020dreaming} for the implementation. During
distillation, the learning rate is set to 0.25, the regularization coefficient
is set to 0.01, and the batch size is set to 84. As illustrated in
\F~\ref{fig:deepinversion}, we employ Jensen-Shannon divergence as the loss
function for back propagation in the white-box setting. However, in the
black-box setting, we employ a heuristic-based search strategy to update the
seed input, as back propagation is \textit{not} possible.

As has illustrated in \F~\ref{fig:distill}, for the white-box setting, it
generates images with notably much higher quality compared to that of the
black-box setting. This illustrates that heuristic-based approaches are hard to
steal knowledge from the victim model, whereas \tool, by successfully
decompiling the ResNet18 executable into high-level model descriptions, enables
gradient-based methods that are much more powerful.
 \end{appendix}

\end{document}